\UseRawInputEncoding
\documentclass[reprint, amsmath, amssymb, aps]{revtex4-1}	
\usepackage{graphicx}
\usepackage{dcolumn}
\usepackage{bm}
\usepackage{mathrsfs}
\usepackage{tikz}
\usepackage{caption}
\usepackage{hyperref}
\input epsf.tex
\usepackage{amssymb}
\usepackage{amsmath}
\usepackage{amsthm}
\usepackage{amsopn}
\usepackage{ulem}
\usepackage{cancel}
\usepackage{color}
\usepackage{float}
\hypersetup{
    colorlinks=true,       
    linkcolor=blue,          
    citecolor=blue,        
    filecolor=magenta,      
    urlcolor=blue           
}

\begin{document}

\title{ Electroweak Phase Transition in an Inert Complex Triplet Model}
\author{MJ. Kazemi}
\email{mj\_kazemi@sbu.ac.ir}
\author{S.S. AbdusSalam}
\email{abdussalam@sbu.ac.ir}
\affiliation{Department of Physics, Shahid Beheshti University, Tehran, Islamic Republic of Iran}

\begin{abstract}

We study the dynamics of electroweak phase transition in a simple extension of the Standard Model where the Higgs sector is extended by adding an $SU(2)_L$-triplet with hypercharge  Y=2. By making random scans over the parameters of the model, we show that there are regions consistent with constraints from collider experiments and the requirement for a strong first-order electroweak phase transition which is needed for electroweak baryogenesis. Further, we also study the power spectrum of the gravitational waves which can be generated due to the first-order phase transitions. Moreover, the detectability of these gravitational waves, via future space-based detectors, is discussed.

\begin{description}
\item[PACS numbers]
03.65.Ta,  05.10.Gg
\end{description}
\end{abstract}
\maketitle

\section{Introduction}\label{sec1}
Cosmological electroweak phase transition (EWPT) is interesting for numerous reasons. It can be a source for primordial magnetic fields \cite{Magnetic}, generate detectable background gravitational waves \cite{GW}, affect the abundance of thermal relic densities for candidate dark matter particles \cite{Wainwright 2009}, and perhaps most importantly, lead to suitable preconditions for baryogenesis \cite{Trodden 1999}.

The Planck experiment collaboration \cite{plank 2016} measured the baryon-to-photon ratio of the universe as $$\frac{n_b}{n_\gamma}=(6.10\pm 0.04)\times 10^{-10}.$$ This is consistent with astronomical measurements of light-element abundances, assuming the standard Big Bang Nucleosynthesis \cite{Trodden 1999, Cyburt 2016}. 
This measured value represents one of the big unresolved particle physics puzzles. It quantifies the matter-antimatter asymmetry of the universe \cite{Trodden 1999, Astro-Baryon-Asymmetry}. The mechanism(s) behind the asymmetry needs to be determined and understood. About half a century ago, A. D. Sakharov proposes three early universe conditions needed to be satisfied for a successful baryon asymmetry generation \cite{Sakharov 1967}: i) baryon number violation, ii) C and CP violation, and iii) departure from thermal equilibrium. These are in principle possible within the framework of electroweak phase transition at the early universe called electroweak baryogenesis (EWBG). The realisation of EWBG within the standard model (SM) of particle physics turned out to be problematic according to lattice simulations \cite{EWPT in SM}. It was found that only cross-over, instead of strongly first-order, phase transitions are possible at the early universe for the observed value of the Higgs boson mass. This indicates that some physics beyond the SM is essential. 

One of the simplest class of models beyond the SM which may lead to strong first order phase transitions and successful EWBG, can be made by  adding an electroweak scalar $SU(2)_L$-multiplet to the SM Higgs sector. A global $Z_2$-symmetry is imposed in constructing this so-called inert multiplet models. With the $Z_2$-symmetry, the lightest neutral component of the new scalar multiplet can be considered as a candidate for dark matter \cite{Cirelli 2006, AbdusSalam 2014}.

The EWPT and its gravitational wave signatures have been well-studied within the framework of the Singlet \cite{Singlet,Alanne:2019bsm,Steele:2013fka,Baum:2020vfl,AbdusSalam:2020uck}, Doublet \cite{Doublet, Huang 2018}, and real Triplet (Y=1) \cite{Huang 2018, Real Triplet, AbdusSalam:2020uck} cases of scalar-multiplet class of models. Another well motivated representation of $SU(2)_L$ group is the complex Triplet (Y=2), which could also be used for explaining the smallness of neutrino mass in Type-II seesaw mechanism \cite{seesawType2,Schechter:1980gr,Schechter:1981cv}.  In this article we address the EWPT of the Inert Complex Triplet model by scanning its parameter space with experimental constraints from Higgs signal strengths imposed. We analyse the parameter space regions that could lead to a first-order EWPT. We also study the power spectra of gravitational waves which could be generated following the first-order transitions. Observation of such  gravitational waves, such as by future space-based gravitational wave detectors \cite{eLISA 2016}, could yield information that is complementary to collider and dark matter experiments.  

This paper is organised as follows. In Sec.~\ref{sec2} we briefly review the Inert Complex Triplet model, describe the relevant parameters for the EWPT, and the theoretical constraints taken into account. In Sec.~\ref{sec3} we describe the results from collider experiments regarding Higgs decay to diphoton and used these to constrain the parameter space of the Inert Complex Triplet model. Finally, in Sec.~\ref{sec4} and Sec.~\ref{sec5} we present the numerical analyses of the model in light of EWPT and gravitational waves generation respectively. 

\section{The model}\label{sec2}
We extend the SM Higgs sector by adding one complex scalar $SU_L(2)$ triplet, $\Delta$, with hypercharge $Y=2$, 
$$
  \Delta=\begin{pmatrix}
    \Delta^{++}\\
    \Delta^{+}\\
    \Delta^{0}\equiv\frac{1}{\sqrt{2}}(S+i A)
  \end{pmatrix}
$$
and impose a $Z_2$ discrete symmetry, under which $\Delta \to -\Delta$ and all other fields unchanged. The most general scalar potential, symmetric under $Z_2$, involving this triplet and the standard $SU_L(2)$ Higgs doublet, $$ H=\begin{pmatrix}
    G^{+}\\
   \frac{1}{\sqrt{2}}(\phi+iG^0)\\
  \end{pmatrix},  
$$ can be written in the following form \cite{AbdusSalam 2014, Hambye 2009},
\begin{eqnarray} 
  V_0=&-& \mu_H^2 H^\dagger H  +\lambda_H (H^\dagger H)^2 \nonumber\\
      &+& \mu_\Delta^2 \Delta^\dagger \Delta + \lambda^{(1)}_\Delta (\Delta^\dagger \Delta)^2+\lambda^{(2)}_{\Delta} |\Delta^\dagger T^a \Delta|^2\nonumber\\
      &+&\lambda^{(1)}_{H\Delta} H^\dagger H \Delta^\dagger \Delta+\lambda^{(2)}_{H\Delta} H^\dagger \tau^a H \Delta^\dagger T^a \Delta. 
\end{eqnarray}
Here, $\tau^a$ and $T^a$  are the $SU(2)$ generators in fundamental and 3's representation respectively. These are normalised such that $\textrm{Tr}[\tau^a , \tau^b]=\frac{1}{2}\delta^{ab}$ and $\textrm{Tr}(T^aT^b)=\frac{1}{2}\delta^{ab}$. Explicitly, $\tau^a=\frac{1}{2}\sigma^a$ where $\sigma^a$s are Pauli matrices and $T^a$s are 
$$ T^1=\frac{1}{\sqrt{2}}\left(
\begin{array}{ccc}
	0 & -1 & 0 \\
	-1 & 0 & 1 \\
	0 & 1 & 0 \\
\end{array}
\right), \  \ 
T^2=\frac{1}{\sqrt{2}}\left(
\begin{array}{ccc}
0 & i & 0 \\
-i & 0 & -i \\
0 & i & 0 \\
\end{array}
\right)
$$
$$
T^3=\left(
\begin{array}{ccc}
1 & 0 & 0 \\
0 & 0 & 0 \\
0 & 0 & -1 \\
\end{array}
\right).$$
We require that $\Delta$ be odd under $Z_2$ symmetry so that the neutral component will not acquire any vacuum expectation value.
There is an electroweak symmetry-breaking minimum at zero temperature, with $\langle H^T\rangle=(0,v/\sqrt{2})$ and $\langle \Delta^T\rangle=(0,0,0)$. In this case, the tree-level field-dependent masses of standard model particles are same as in the SM,  
$$ m^2_t(\phi)=\frac{y_t^2}{2}\phi^2, \ \ \ \ \  m^2_b(\phi)=\frac{y_b^2}{2}\phi^2,$$
$$ m^2_W(\phi)=\frac{g^2}{4}\phi^2, \ \ \ \ \  m^2_Z(\phi)=\frac{g^2+g'^2}{4}\phi^2,$$
$$ m_{h}^2(\phi)=-\mu_H^2+3\lambda_H \phi^2,$$
and the masses of the component of the additional Triplet scalar are given by 
$$
m_{S}^2(\phi)=m_{A}^2(\phi)=\mu_\Delta^2+\frac{1}{2}(\lambda^{(1)}_{H\Delta}+\frac{1}{2} \lambda^{(2)}_{H\Delta}) \phi^2,
$$
$$
m_{\Delta^{+}}^2(\phi)=\mu_\Delta^2+\frac{1}{2}\lambda^{(1)}_{H\Delta}\phi^2,
$$
$$
m_{\Delta^{++}}^2(\phi)=\mu_\Delta^2+\frac{1}{2}(\lambda^{(1)}_{H\Delta}-\frac{1}{2} \lambda^{(2)}_{H\Delta}) \phi^2.
$$
This model has five real parameters in addition to those of the SM. However, only three of them, i.e. the triplet mass parameter and the doublet-triplet couplings, appear in the tree-level triplet masses. Thus only $\mu_{\Delta}$, $\lambda^{(1)}_{H\Delta}$ and $\lambda^{(2)}_{H\Delta}$ parameters are relevant for electroweak phase transition dynamics, at one-loop approximation (see Sec.~\ref{sec3}).

In the next section, we first study some  constrains which come from collider phenomenology on these parameters (or equivalently on the mass spectrum of the Triplet scalar at $\left< \phi \right > = v = 246 \textrm{ GeV}$). After this, we will study the parameter space to find regions that can lead to (i) strong first order phase transitions, and (ii) detectable gravitational waves. While sampling the parameter space, we apply theoretical constraints, checking that \textit{unitarity} and \textit{vacuum stability} conditions on the triplet self-couplings are satisfied before applying those from the Higgs signal strength measurements. For a stable vacuum, the scalar potential should be bounded from below along all possible field directions. At the tree level, the vacuum stability requirement leads to \cite{AbdusSalam 2014},
\begin{eqnarray}
  &&\lambda_H, \lambda_{\Delta}^{(1)},\lambda_{\Delta}^{(2)}>0\nonumber\\
  && -2\sqrt{\lambda_H( \lambda_{\Delta}^{(1)}+\lambda_{\Delta}^{(2)})}<2\lambda_{H\Delta}^{(1)}+\lambda_{H\Delta}^{(2)}  \nonumber\\
  && -2\sqrt{\lambda_H \lambda_{\Delta}^{(1)}}< \lambda_{H\Delta}^{(1)}   \nonumber\\
  && -2\sqrt{\lambda_H( \lambda_{\Delta}^{(1)}+\lambda_{\Delta}^{(2)})}<2\lambda_{H\Delta}^{(1)}-\lambda_{H\Delta}^{(2)}. \nonumber 
\end{eqnarray}

In what follows, we begin by considering the implications of $h\rightarrow \gamma\gamma$ and $h\rightarrow Z\gamma$ limits on the parameter regions of the Inert Complex Triplet model which from now on we address as the inert triplet model (ITM). 

\section{excluding parameter space via Higgs decay rates}\label{sec3}
The branching ratios of the Higgs decays in the ITM differ from the SM ones. As such the Higgs decay measurement or limits can be used as a probe for ITM. Specifically, the Higgs-to-diphoton channel, $h \to \gamma\gamma$, because of its relatively clean signature at the Large Hadron Collider (LHC), could play an important role for this purpose. Here we analyse the ITM parameter space by using recent ATLAS and CMS results for the Higgs-to-diphoton signal strength. We find that a significantly large region of the parameter space are excluded via these recent data. 

To study the ITM contributions to $h \to \gamma\gamma$ decay rate, we address the ratio, 
\begin{align}\label{rgg}
R_{\gamma \gamma}&\equiv \frac{\sigma(pp\to h\to \gamma\gamma)^{\textrm{ITM}}}{\sigma(pp\to h\to \gamma\gamma)^{\textrm  {SM}}}\nonumber\\*
&\approx\frac{\left[\sigma(gg\to h) \textrm{Br}(h\to\gamma\gamma)\right]^{\textrm {ITM}}}{\left[\sigma(gg\to h) \textrm{Br}(h\to\gamma\gamma)\right]^{\textrm {SM}}}.
\end{align}
Here the fact that the gluon-gluon fusion is the dominant channel for Higgs production were used. Moreover, since $\sigma(gg\to h)$ is the same in both  the ITM and SM, the $R_{\gamma \gamma}$ reduces to \cite{Arhrib 2012}
\begin{equation}
R_{\gamma \gamma}=\frac{\Gamma_{h}^{\textrm {SM}}}{\Gamma_{h}^{\textrm {ITM}}}\frac{\Gamma(h\to\gamma\gamma)^{\textrm {ITM}}}{\Gamma(h\to\gamma\gamma)^{\textrm {SM}}}.
\end{equation}
In similar way, for $Z\gamma$ decay channel, an analogous quantity, $R_{Z \gamma}$, can be defined as 
\begin{equation}
R_{Z \gamma}=\frac{\Gamma_{h}^{\textrm {SM}}}{\Gamma_{h}^{\textrm {ITM}}}\frac{\Gamma(h\to Z\gamma)^{\textrm {ITM}}}{\Gamma(h\to Z\gamma)^{\textrm {SM}}}.
\end{equation}

Within the SM, many channels contribute to the total decay width of the Higgs boson. The most important ones for $m_h=125 \textrm{ GeV}$ are $b\overline{b}$, $c\overline{c}$, $\tau^+\tau^-$, $ZZ^*$, $WW^*$, $\gamma\gamma$, $Z\gamma$ and $gg$. Hence the total Higgs decay widths is approximately given by:
\begin{eqnarray}
\Gamma_{h}^{\textrm{SM}}&=&\sum_{f=\tau,b,c}\Gamma_{h\to f\bar{f}}^{\textrm{SM}} + \Gamma_{h\to WW^*}^{\textrm{SM}}\nonumber\\
&+&\Gamma_{h\to ZZ^*}^{\textrm{SM}} + \Gamma_{h\to gg}^{\textrm{SM}}+\Gamma_{h\to \gamma\gamma}^{\textrm{SM}}+\Gamma_{h\to Z\gamma}^{\textrm{SM}}\nonumber.
\end{eqnarray}
In the ITM, the total decay width of the Higgs can be modified with respect to the SM, since the charged scalars exchanged in loops give extra contributions to the $h\to\gamma\gamma$ and $h\to Z\gamma$ amplitudes \cite{Gunion 2000}. 
In addition, the total decay width  changes due to the existence of additional decay channels, i.e.  $h\to SS$, $h\to AA$, $h\to \Delta^{\pm}\Delta^{\mp}$ and $h\to \Delta^{\pm\pm}\Delta^{\mp\mp}$ \cite{Arhrib 2012, Swiezewska 2013}. So 
\begin{eqnarray}
\Gamma_{h}^{\textrm{ITM}}&=&\sum_{f=\tau,b,c}\Gamma_{h\to f\bar{f}}^{\textrm{ITM}} + \Gamma_{h\to WW^*}^{\textrm{ITM}}\nonumber\\
&+&\Gamma_{h\to ZZ^*}^{\textrm{ITM}} + \Gamma_{h\to gg}^{\textrm{ITM}}+\Gamma_{h\to \gamma\gamma}^{\textrm{ITM}}+\Gamma_{h\to Z\gamma}^{\textrm{ITM}}\nonumber \\
&+&\Gamma_{h\to AA}^{\textrm{ITM}} + \Gamma_{h\to SS}^{\textrm{ITM}}+\Gamma_{h\to \Delta^{+}\Delta^{-}}^{\textrm{ITM}}+\Gamma_{h\to \Delta^{++}\Delta^{--}}^{\textrm{ITM}}.\nonumber
\end{eqnarray}
The decay rates of these additional decay channels, when they are kinematically open, $2 m_\varphi< m_h$, are given by
\begin{equation}
\Gamma_{h\to \varphi\varphi^{\dag}}^{\textrm{ITM}}=\frac{\xi_\varphi \lambda_{h\varphi\varphi^{\dag}}^2}{8\pi m_h}\sqrt{1-\frac{4m_{\varphi}^2}{m_h^2}},
\end{equation}
where $\lambda_{h\varphi\varphi^{\dag}}=(m_{\varphi}^2-\mu_{\Delta}^2)/v$ and $\xi_\varphi=1$ for $\varphi=A, \ S$ and $\lambda_{h\varphi\varphi^{\dag}}=2(m_{\varphi}^2-\mu_{\Delta}^2)/v$ and $\xi_\varphi=1/2$ for charged scalars, $\varphi=\Delta^{+}, \ \Delta^{++}$.

The partial widths of the tree-level Higgs decays into SM particles, and  the loop-mediated decay into $gg$ in the ITM are equal to the corresponding ones in the SM; for completeness they are summarised in Appendix A. The $h\to Z\gamma$ and $h\to\gamma\gamma$ SM processes get modified within the ITM. In SM these decays are dominated by contributions from the $W$ gauge boson and top quark loops while in the ITM, the couplings of the Higgs doublet to the triplet scalars modify these decays via the following one-loop diagrams. 
\begin{figure}[H]
\begin{tikzpicture}
\draw (0,0) node[above right]{\includegraphics[width=.95\linewidth, trim={0cm 1.5cm 0cm 1.5cm}]{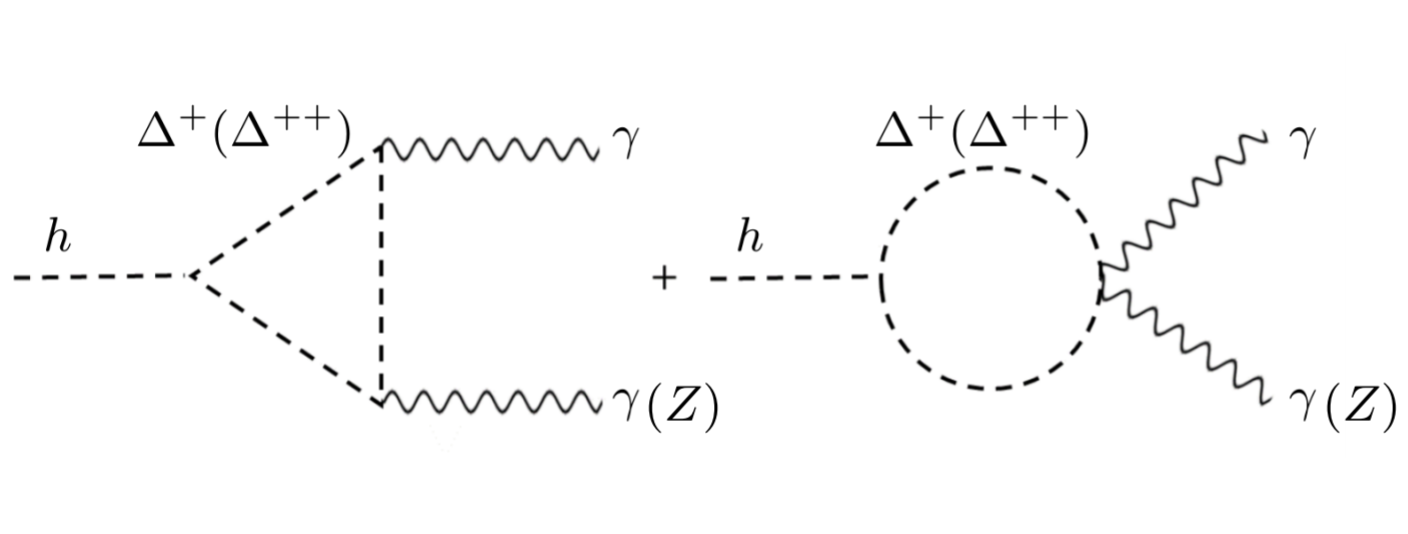}};
\end{tikzpicture}
\caption{Feynman diagrams for charged scalar particles contributing to $h\to \gamma\gamma$ or  $h\to Z\gamma$.}\label{fig1}
\end{figure}

Following the general results for spin-0, spin-1/2 and spin-1 contributions to these decay rates \cite{Gunion 2000, Chen 2013} which can be obtained using the Feynman rules listed in Ref.~\cite{Denner 1993}, the modified decays in ITM are given by \cite{Chen-2 2013}
\begin{eqnarray}
\Gamma_{H\rightarrow\gamma\gamma}^\text{ITM}&=&\frac{\alpha^2}{256 \pi^3 v^2}m_h^3\left| {\cal A}_\mathrm{ITM}^{\gamma\gamma}\right|^2, \textrm{ and }
\end{eqnarray}
\begin{eqnarray}
\Gamma_{h\to Z\gamma}^{\text{ITM}}&=& \frac{\alpha^2}{512\pi^3 v^2}m_h^3\left(1-\frac{m_Z^2}{m_h^2}\right)^3 \left|{\cal A}_\mathrm{ITM}^{Z\gamma}\right|^2. 
\end{eqnarray}
\begin{equation}
\textrm{ Here, } {\cal A}_\mathrm{SM}^{\gamma\gamma}=A_1^{\gamma\gamma}\big(\tau_w)+\frac{4}{3} A_{\frac{1}{2}}^{\gamma\gamma}\big(\tau_t\big)\nonumber, 
\end{equation}
\begin{eqnarray}
{\cal A}_\mathrm{ITM}^{\gamma\gamma}={\cal A}_\mathrm{SM}^{\gamma\gamma}+g_{\Delta^{+}}^{\gamma\gamma} A_0^{\gamma\gamma}(\tau_{\Delta^{+}})+g_{\Delta^{++}}^{\gamma\gamma} A_0^{\gamma\gamma}(\tau_{\Delta^{++}})\nonumber, 
\end{eqnarray}
\begin{eqnarray}
 {\cal A}_\mathrm{SM}^{Z\gamma}= 2\frac{c_w}{s_w} A_1^{Z\gamma}(\tau_W,\lambda_W)
+4\frac{(1-\frac{8}{3} s_w^2)}{s_w c_w}A_{1/2}^{Z\gamma}(\tau_t,\lambda_t)\nonumber,
\end{eqnarray}
\begin{eqnarray}
{\cal A}_\mathrm{ITM}^{Z\gamma}&=&{\cal A}_\mathrm{SM}^{Z\gamma}- g_{\Delta^{+}}^{Z\gamma} A_0^{Z\gamma}(\tau_{\Delta^{+}},\lambda_{\Delta^{+}})- g_{\Delta^{++}}^{Z\gamma} A_0^{Z\gamma}(\tau_{\Delta^{++}},\lambda_{\Delta^{++}})\nonumber, 
\end{eqnarray}
\\
$\tau_i=4m_i^2/m_h^2$, $\lambda_i=4m_i^2/m_Z^2$ ($i=W,t,\Delta^{+},\Delta^{++}$), $s_w=\sin \theta_w$, and $c_w=\cos \theta_w$. $\theta_w$ is the Weinberg mixing angle and the coupling constants are given by 
\begin{eqnarray}
g_{\Delta^{+}}^{\gamma\gamma}&=&\frac{m_{\Delta^{+}}^2-\mu_{\Delta}^2}{m_{\Delta^{+}}^2},\nonumber\\
g_{\Delta^{++}}^{\gamma\gamma}&=&\frac{4\left(m_{\Delta^{++}}^2-\mu_{\Delta}^2\right)}{m_{\Delta^{++}}^2},\nonumber\\
g_{\Delta^{+}}^{Z\gamma}&=&\frac{4(m_{\Delta^{+}}^2-\mu_{\Delta}^2)(-s_w^2)}{ m_{\Delta^{+}}^2 s_w c_w },\nonumber \textrm{ and }\\
g_{\Delta^{++}}^{Z\gamma}&=&\frac{8(m_{\Delta^{++}}^2-\mu_{\Delta}^2)(1-2s_w^2)}{ m_{\Delta^{++}}^2 s_w c_w }\nonumber . 
\end{eqnarray}
The loop functions $A^{\gamma\gamma}_{(0,\,1/2,\,1)}$ and $A^{Z\gamma}_{(0,\,1/2,\,1)}$ are given in Appendix B.

\begin{widetext}

\begin{figure}[H]
\centering
\includegraphics[width=1\linewidth, trim={0cm 0cm 0cm 0cm}]{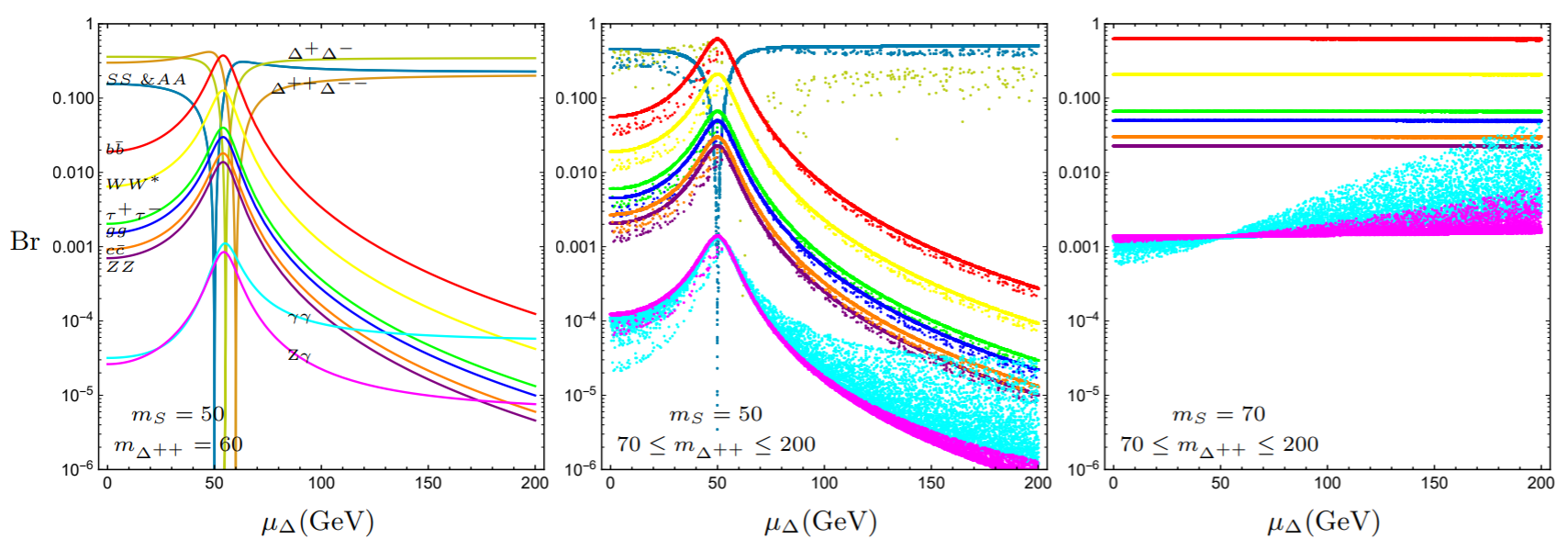}
\caption{Branching ratios for $h$ with mass 125 GeV. Left panel: decay channels $h\to AA$, $h\to SS$, $h\to \Delta^{+}\Delta^{-}$ and $h\to \Delta^{++}\Delta^{--}$ are open  ($m_S=m_A=50$ GeV, $m_{\Delta^{++}}=60$ GeV). Middle panel: $h\to \Delta^{+}\Delta^{-}$ and $h\to \Delta^{++}\Delta^{--}$ are open. Right panel: no $h$ decay channels to triplet particles allowed ($m_S=70$ GeV, $m_{\Delta^{++}}>m_{\Delta^{+}}>70$ GeV).}\label{fig2}
\end{figure}
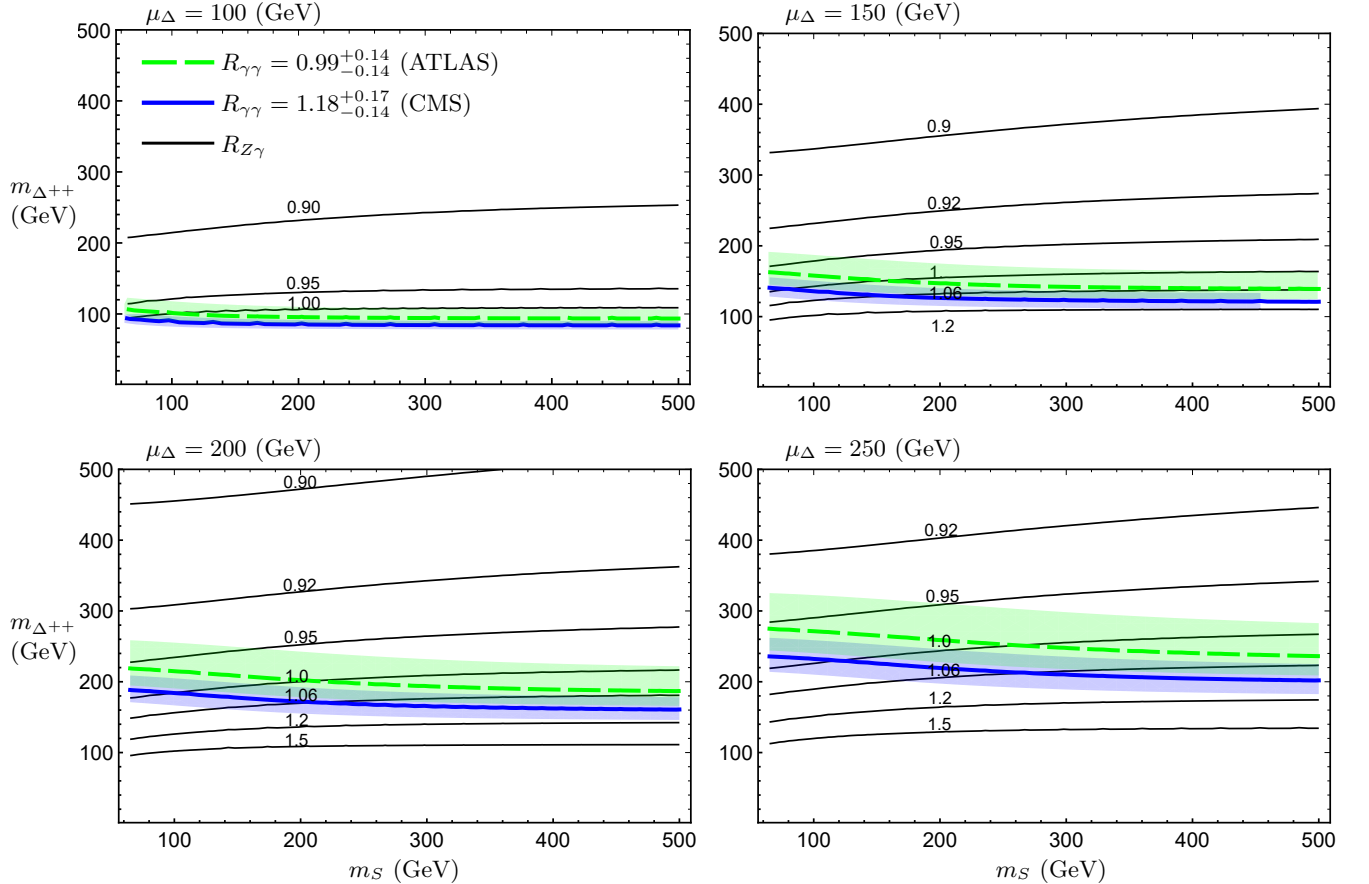
\begin{figure}[H]
\centering
\begin{tikzpicture}
\draw (-.9,.2) node[above right]{\includegraphics[width=.46\linewidth, trim={0cm 0cm 0cm 0cm}]{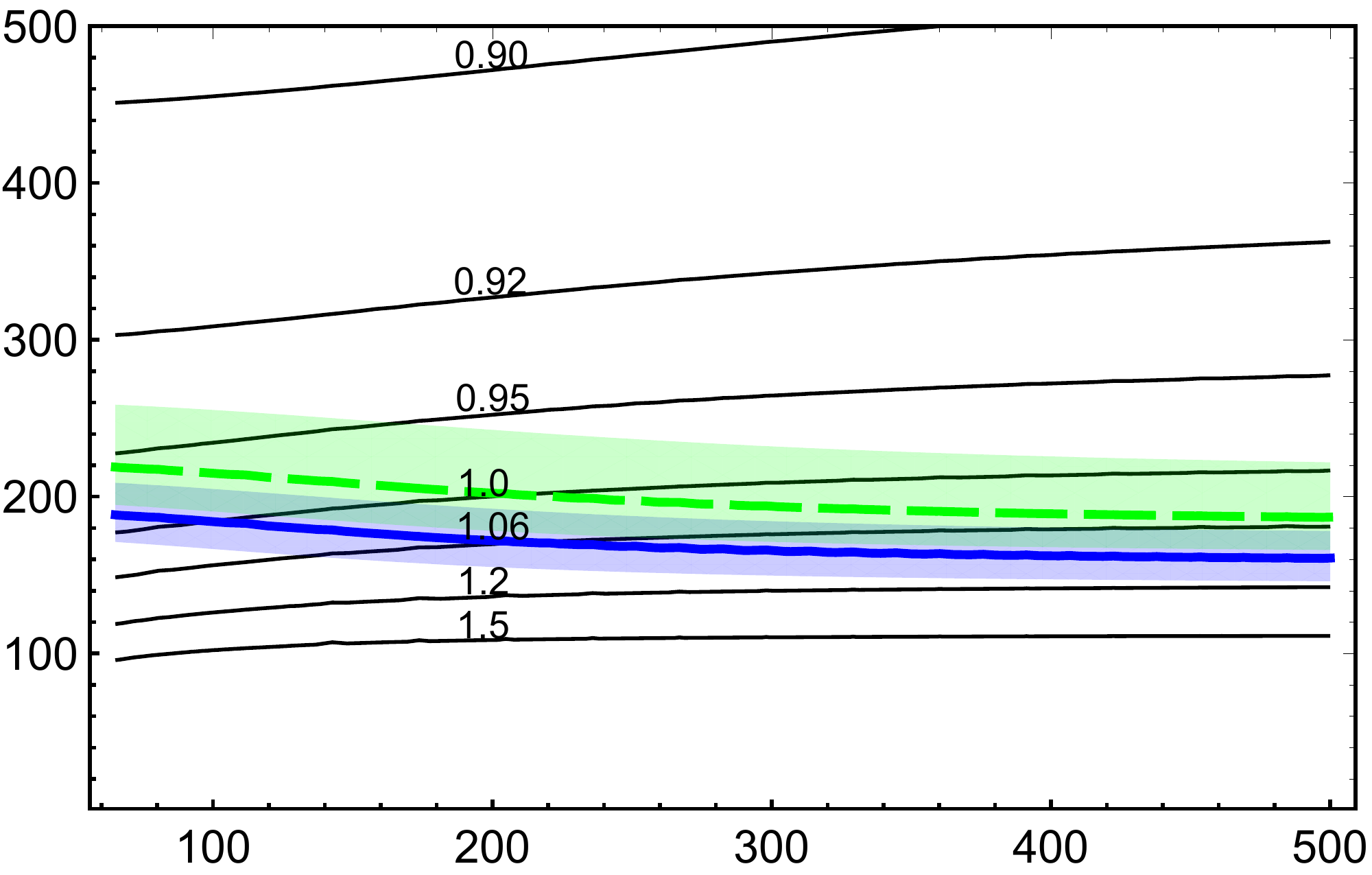}};
\draw (7.6,.2) node[above right]{\includegraphics[width=.46\linewidth, trim={0cm 0cm 0cm 0cm}]{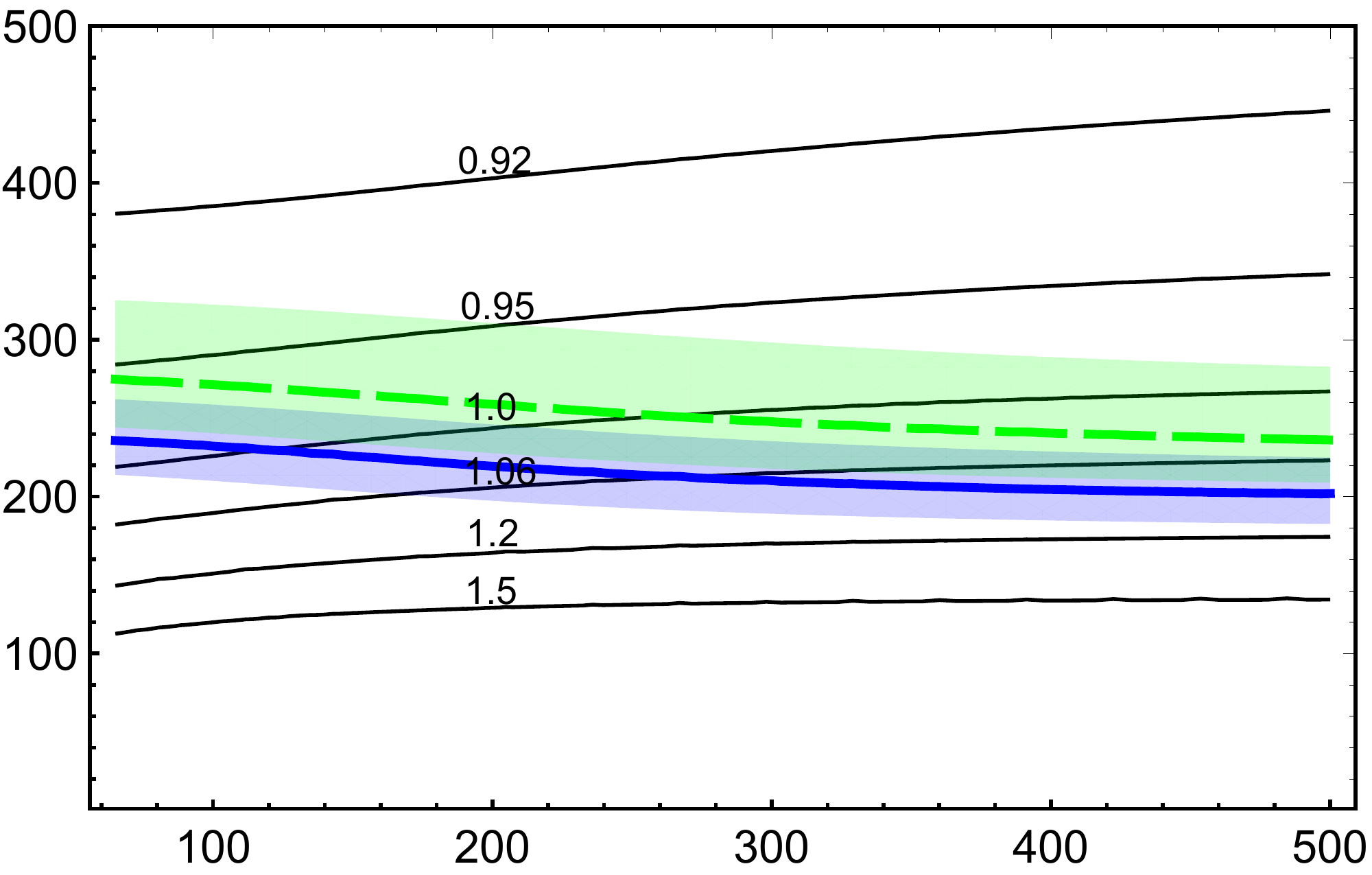}};
\draw (-.9,6) node[above right]{\includegraphics[width=.46\linewidth, trim={0cm 0cm 0cm 0cm}]{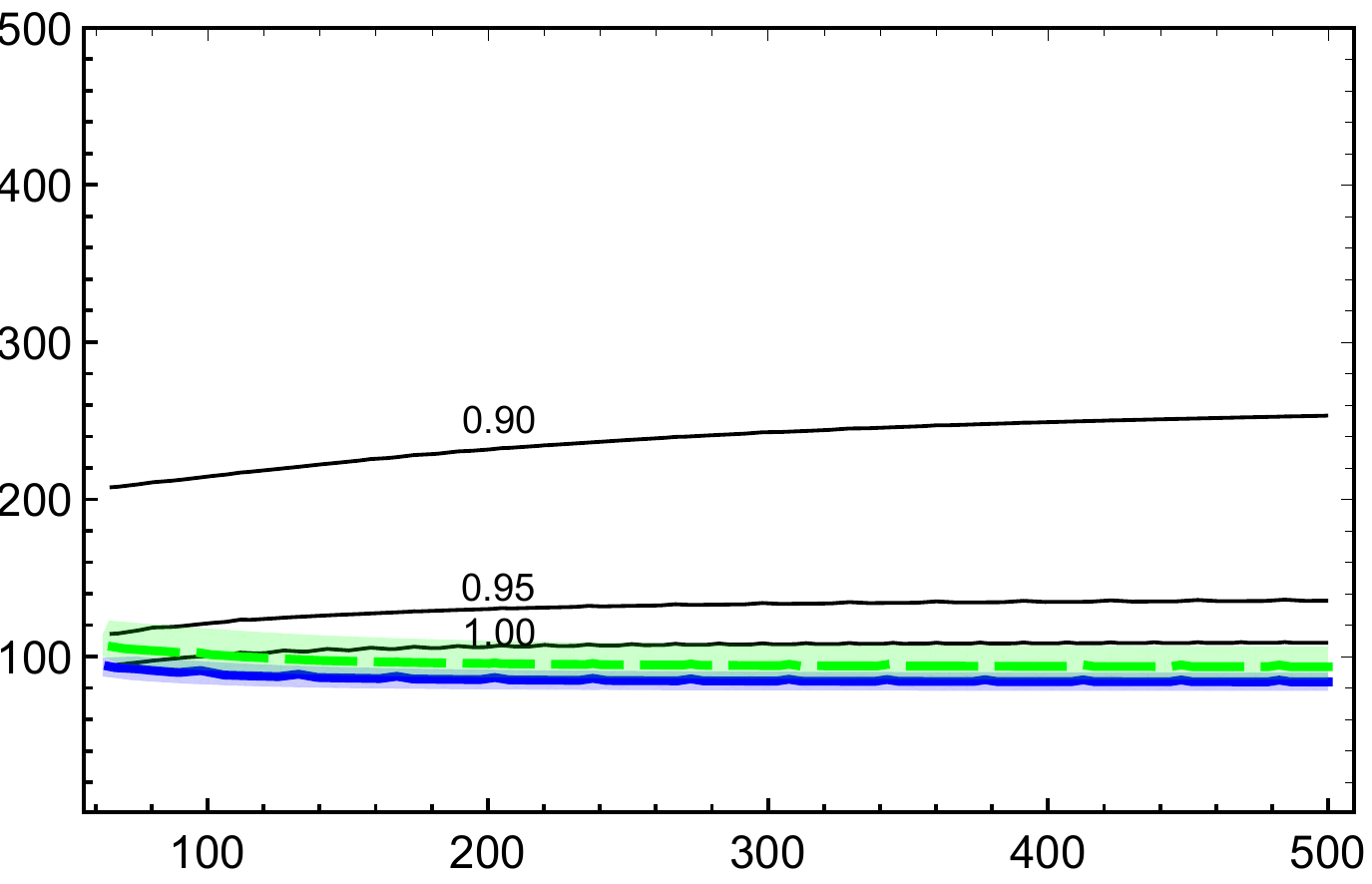}};
\draw (7.6,6) node[above right]{\includegraphics[width=.46\linewidth, trim={0cm 0cm 0cm 0cm}]{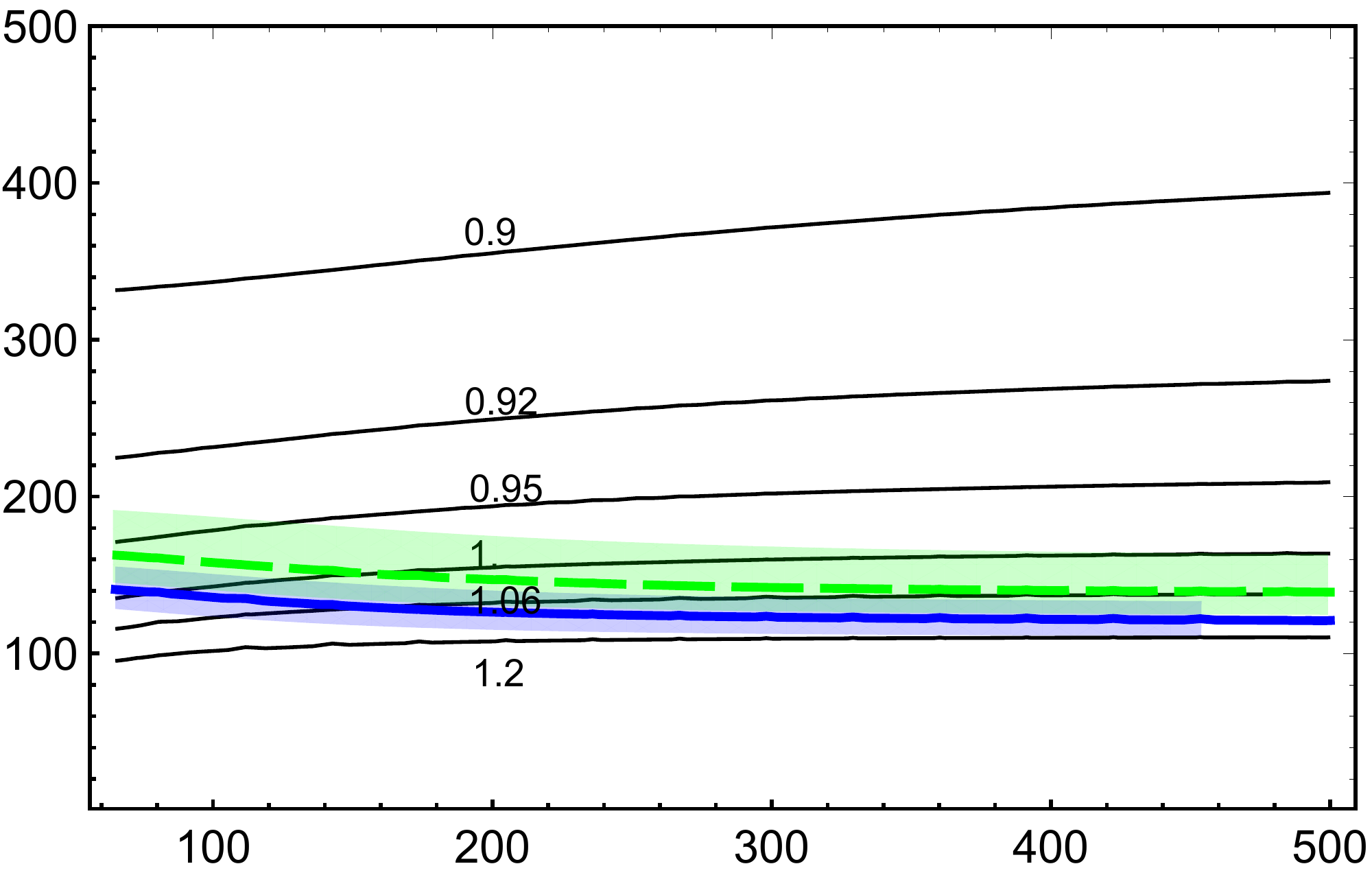}};
\draw (1,10.5) node[above right]{$R_{\gamma\gamma}=0.99^{+0.14}_{-0.14}$ (ATLAS)};
\draw (1,9.95) node[above right]{$R_{\gamma\gamma}=1.18^{+0.17}_{-0.14}$ (CMS)};
\draw (1,9.44) node[above right]{$R_{Z\gamma}$};
\draw (-.5,9.3) node[above right]{\includegraphics[width=.1\linewidth, trim={0cm 0cm 0cm 0cm}]{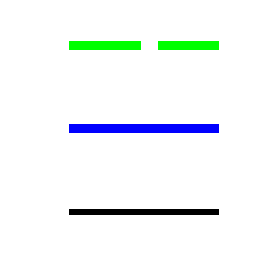}};
\draw (-1.8,3.1) node[above right]{$m_{\Delta^{++}}$};
\draw (-1.8,2.7) node[above right]{(GeV)};
\draw (-1.8,8.9) node[above right]{$m_{\Delta^{++}}$};
\draw (-1.8,8.5) node[above right]{(GeV)};
\draw (2.7,-.2) node[above right]{$m_{S}$ (GeV)};
\draw (11.3,-.2) node[above right]{$m_{S}$ (GeV)};
\draw (0,11.2) node[above right]{$\mu_\Delta=100$ (GeV)};
\draw (8.5,11.2) node[above right]{$\mu_\Delta=150$ (GeV)};
\draw (0,5.4) node[above right]{$\mu_\Delta=200$ (GeV)};
\draw (8.5,5.4) node[above right]{$\mu_\Delta=250$ (GeV)};
\end{tikzpicture}
\caption{The regions of triplet mass spectrum which are consistent with the measured values of $R_{\gamma\gamma}$ in ATLAS and  CMS, for $\mu_\Delta$ equals to 100, 150, 200 and 250 (GeV). The solid black contour lines represent $R_{Z\gamma}$. The blues and green bands capture the 1 $\sigma$ uncertainties for the ATLAS and CMS results. The light-blue band is for the CMS' observed value of $R_{\gamma\gamma}$ whose central value is represented by the solid blue line. The light-green band is for the ATLAS' observed value of $R_{\gamma\gamma}$ whose central value is represented by the broken green line.}\label{fig3}
\end{figure}

\end{widetext}
In the Fig.\ref{fig2} we plot the branching rations for some regions of parameter space. It turns out that when the triplet decay channels, $h\to AA$, $h\to SS$, $h\to \Delta^{+}\Delta^{-}$ and $h\to \Delta^{++}\Delta^{--}$ are kinematically allowed, their partial widths dominate over the  partial widths of decays into SM particles. Therefore, in this case, the value of $R_{\gamma\gamma}$ deviates significantly from $R_{\gamma\gamma} = 1$. This is not consistent with experimental results. Conversely, when these decay channels are kinematically closed, ($m_A, m_S, m_{\Delta^{+}}, m_{\Delta^{++}}>m_h/2$), the total width of $h$ is slightly modified with respect to the SM case, since the branching ratios of $h\to \gamma \gamma$ and $h\to Z \gamma$, which are the only processes that receive contributions from triplet scalars, are of the order of $10^{-2}$.

For the numerical analysis, we scan the parameter space of the ITM in the range 
$10 \ \textrm{GeV}\leq m_S , m_{\Delta^{++}}\leq 500\ \textrm{GeV}$,
for some specific values of $\mu_\Delta$; 100, 150 , 200, 250 (GeV). We then compare the values of $R_{\gamma\gamma}$ obtained with the most recent measurements by the ATLAS \cite{ATLAS 2018} and CMS \cite{CMS 2018} collaborations: 
$$R^{\textbf{ATLAS}}_{\gamma\gamma}= 0.99 \pm 0.14  \textrm{ and } R^{\textbf{CMS}}_{\gamma\gamma}=1.18^{+0.17}_{-0.14}.$$ 
In fact we have found that the $R_{\gamma\gamma}$ enhancement is only possible when $m_S, m_{\Delta^{++}}>m_h/2$. In Fig.\ref{fig3} we illustrate the regions of the parameter space allowed by these experimental constraints. The coloured bands represented the regions within the reported experimental uncertainties. We also superimpose the contour lines which represent the values of $R_{Z\gamma}$. The decay $h\to Z\gamma$ has not been discovered \cite{Goertz 2020}, but $R_{Z\gamma}$ is rather bounded from above to be less than 3.6 at 95\% C.L. \cite{ATLAS 2020}.




In the next section we discuss how these experimental results can constrain the properties of electroweak phase transitions and the gravitational wave spectra that could follow. 

\section{Dynamics of the EWPT}\label{sec4}
In order to study the electroweak phase transition, we need to follow the evolution of the Higgs vacuum expectation value, i.e. the minimum of the Higgs effective potential over the thermal history of the universe. For this, we use the standard techniques of finite temperature field theory \cite{Quiros 1999,Croon:2020cgk}. The one-loop level Higgs effective potential at finite temperatures can be written as 
\begin{equation}
V(\phi,T)=V_0(\phi)+V_{CW}(\phi)+V_T(\phi,T)
\end{equation}
where the tree-level potential is given by
$$V_0(\phi)=-\frac{1}{2}\mu_H \phi^2+\frac{1}{4}\lambda_H \phi^4.$$
The zero- and finite-temperature corrections at one-loop, i.e. the Coleman-Weinberg potential $V_{CW} (\phi)$  \cite{footnote1,Coleman 1973} and $V_{T}(\phi,T)$ are respectively given by \cite{Parwani 1992, Weinberg 1987}
$$ V_{CW}=\sum_i g_i\left[m_i^4(\phi)\left(\log\frac{m_i^2(\phi)}{m_i^2(v_0)}-\frac{3}{2}\right)+2m_i^2(\phi)m_i^2(v_0)\right],$$
and
$$V_{T}=\sum_i 32 g_iT\int_0^\infty  dxx^2\ln\left[1-(-1)^{F_i} e^{-\sqrt{x^2+m_i^2(\phi)}/T}\right].$$
Here $i=\{W,Z,t,b,h, A, S, \Delta^{+},\Delta^{++}\}$, $F_i$ represents the fermionic number, $n_i$ is the number of degrees of freedom of the different species of particles, $$n_i=\{6,3,12,12,1, 1, 1, 2, 2\},$$ and  $g_i\equiv(-1)^{F_i}n_i/64\pi^2$. 

A first-order phase transition happens when the effective potential has two minima of the same value at some critical temperature, $T_c$. In such case the system can transit between the vacua via thermal fluctuations or quantum tunnelling. This transition physically means creation of spherically symmetric regions of \textit{true vacuum}, bubbles of the broken phase, expanding in the background of the \textit{false vacuum}.

\begin{widetext}

\begin{figure}[H]
\centering
\begin{tikzpicture}
\draw (-.9,.2) node[above right]{\includegraphics[width=.46\linewidth, trim={0cm 0cm 0cm 0cm}]{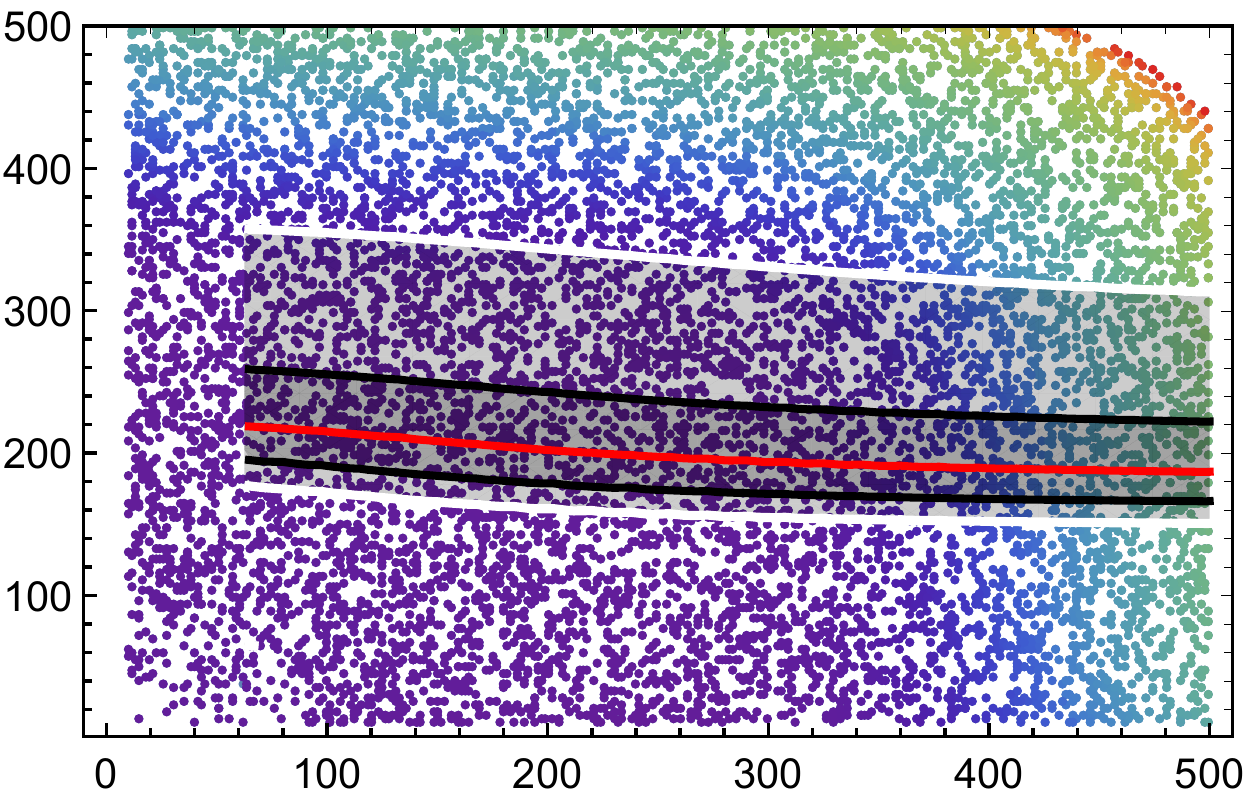}};
\draw (7.6,.2) node[above right]{\includegraphics[width=.46\linewidth, trim={0cm 0cm 0cm 0cm}]{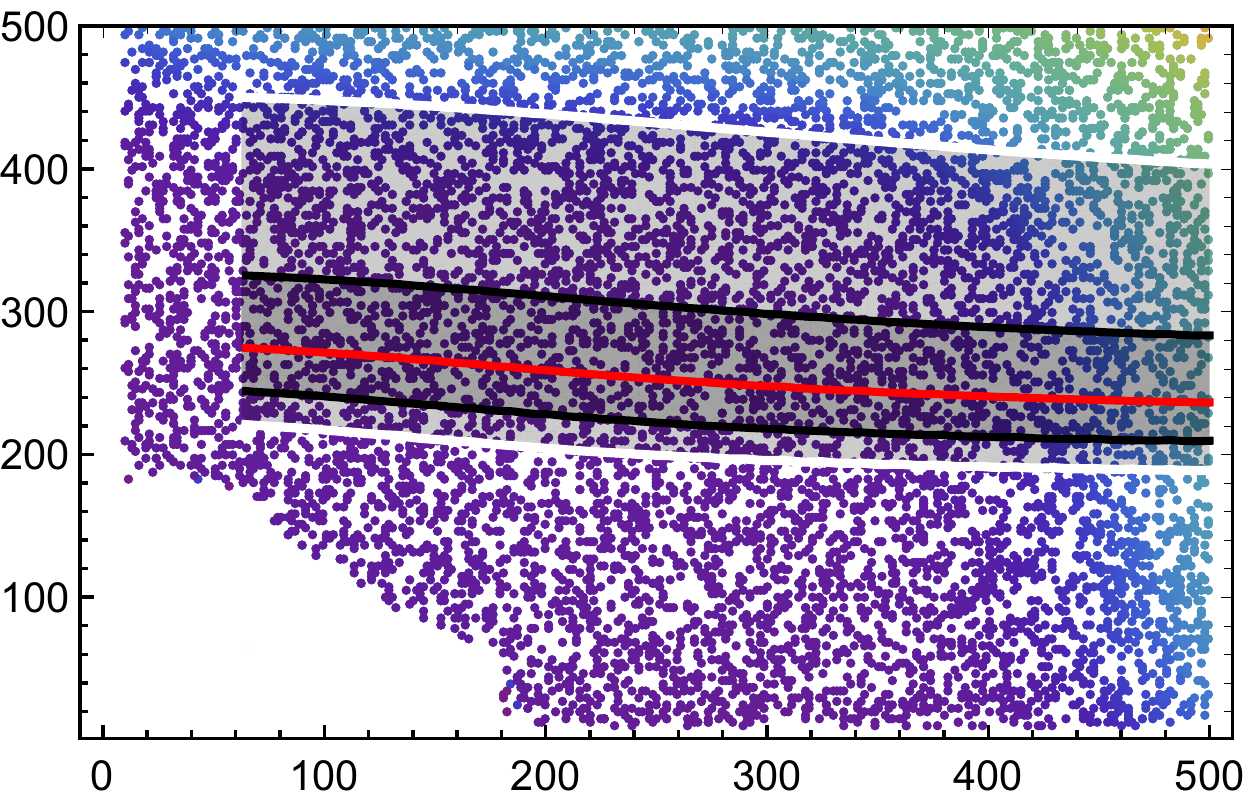}};
\draw (-.9,6) node[above right]{\includegraphics[width=.46\linewidth, trim={0cm 0cm 0cm 0cm}]{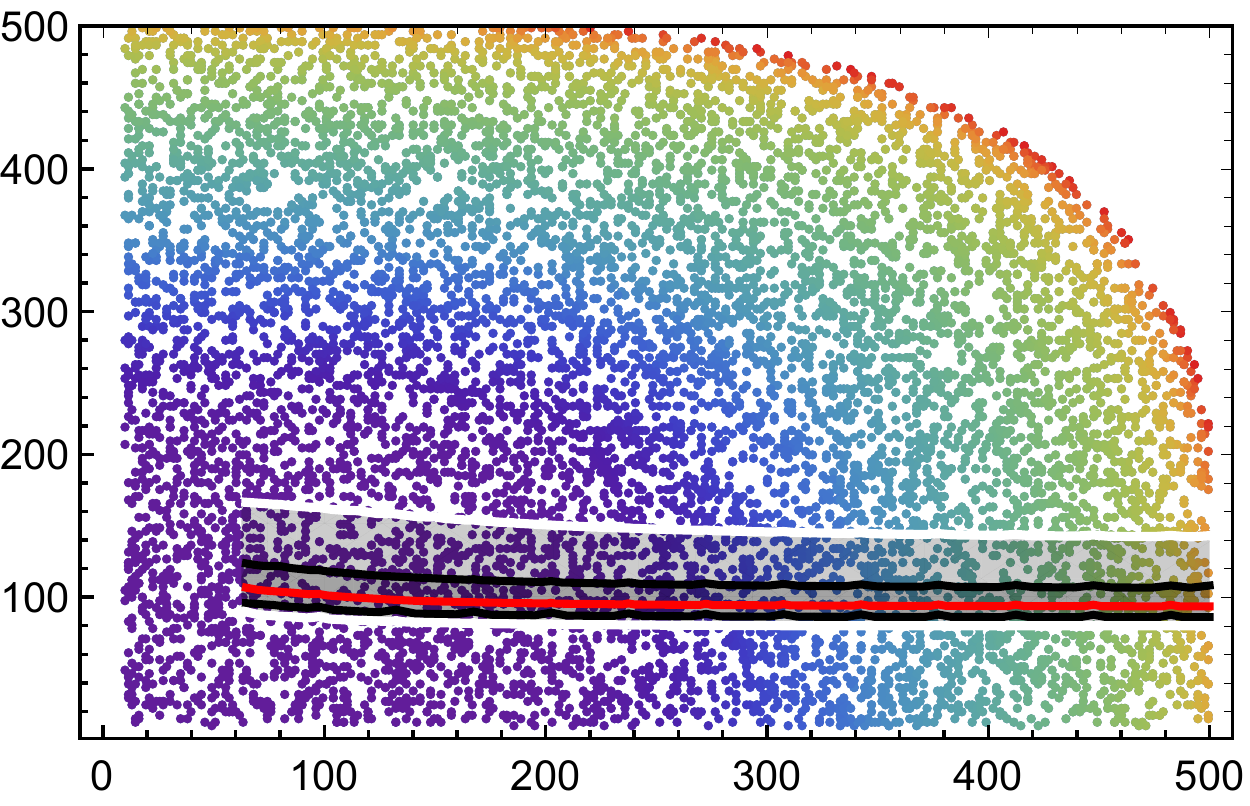}};
\draw (7.6,6) node[above right]{\includegraphics[width=.46\linewidth, trim={0cm 0cm 0cm 0cm}]{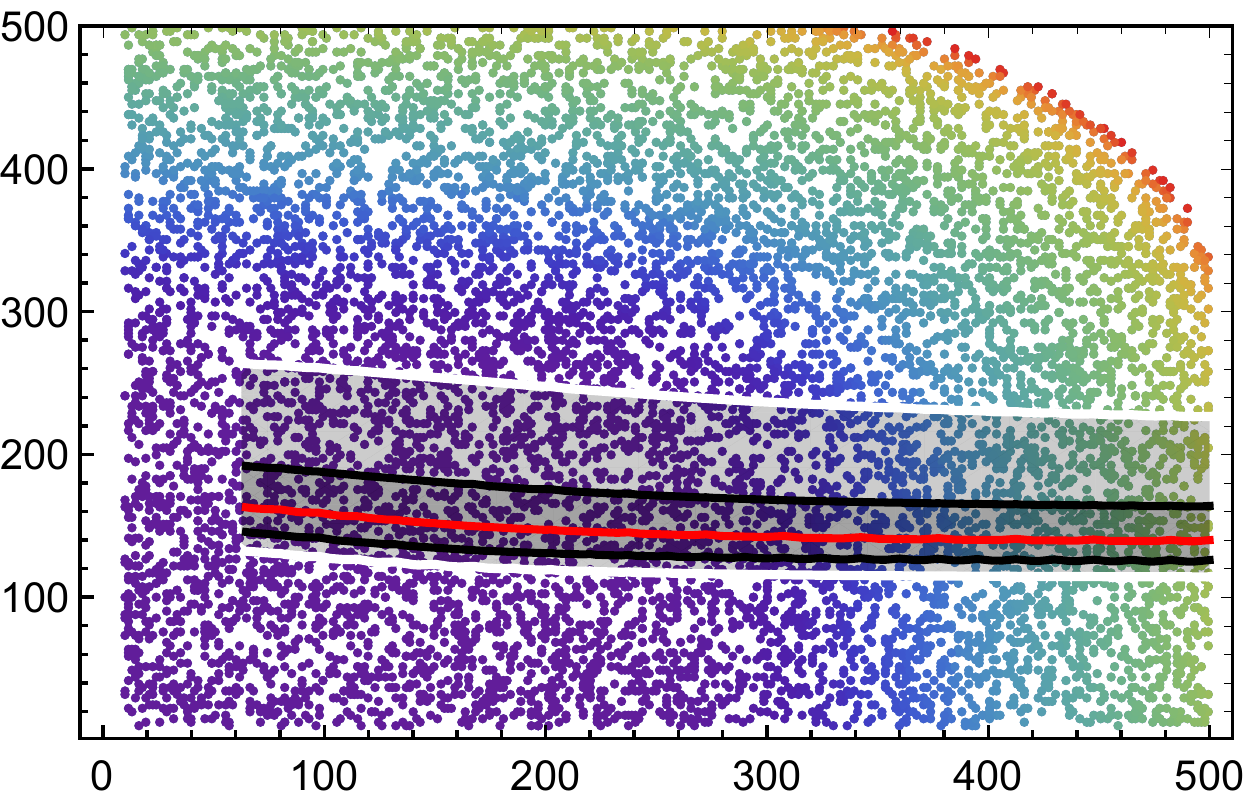}};
\draw (15.7,5.91) node[above right]{\includegraphics[width=.0429\linewidth, trim={0cm 0cm 0cm 0cm}]{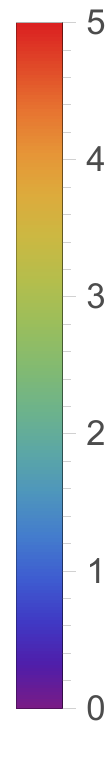}};
\draw (15.7,.1) node[above right]{\includegraphics[width=.0429\linewidth, trim={0cm 0cm 0cm 0cm}]{BarLegend.pdf}};
\draw (-1.8,3.1) node[above right]{$m_{\Delta^{++}}$};
\draw (-1.9,2.7) node[above right]{(GeV)};
\draw (-1.8,8.9) node[above right]{$m_{\Delta^{++}}$};
\draw (-1.9,8.5) node[above right]{(GeV)};
\draw (2.7,-.2) node[above right]{$m_{S}$ (GeV)};
\draw (11.2,-.2) node[above right]{$m_{S}$ (GeV)};
\draw (15.6,5.5) node[above right]{$\phi_c/T_c$};
\draw (15.6,11.3) node[above right]{$\phi_c/T_c$};
\draw (0,11.2) node[above right]{$\mu_\Delta=100$ (GeV)};
\draw (8.5,11.2) node[above right]{$\mu_\Delta=150$ (GeV)};
\draw (0,5.4) node[above right]{$\mu_\Delta=200$ (GeV)};
\draw (8.5,5.4) node[above right]{$\mu_\Delta=250$ (GeV)};
\end{tikzpicture}
\caption{The coloured points represent the region with first-order electroweak phase transition. The colour of points demonstrate the strength of the phase transition, $\phi_c/T_c$, at the critical temperature. The red lines show parameter regions which are compatible with the ATLAS result for $R_{\gamma\gamma}$. The black (white) lines represent the boundary of 1 $\sigma$ (2 $\sigma$) bands for the ATLAS result.}\label{fig4}
\end{figure}

\end{widetext}

In the standard EWBG scenario, the SM fermions interact with the bubble walls in a CP-violating manner. This leads to a chiral asymmetry production in front of the bubble wall which can subsequently turn to baryon generation via \textit{sphaleron processes} which convert the chiral asymmetry to baryon asymmetry  \cite{Trodden 1999, Cline 2018}. The generated baryons could then fall into the growing bubble. This EWBG mechanism could work for explaining matter-antimatter asymmetry of the universe if the generated baryon asymmetry is not washed out by sphalerons inside the bubble \cite{Trodden 1999, Cline 2018}. This condition requires that \cite{Trodden 1999, Moore 1998}
\begin{equation}
\frac{\phi_c}{T_c} > 1
\end{equation}
where $\phi_c$ is the Higgs vacuum expectation value 
at the critical temperature, $T_c$. We numerically check this condition within the ITM by randomly scanning the ITM parameters allowed within the ranges 
$$10 \ \textrm{GeV}\leq m_S , m_{\Delta^{++}}\leq 500\ \textrm{GeV,}$$
for fixed values of $\mu_{\Delta} = 100,\ 150,\ 200$ and $250$ (GeV). The results are shown in Fig.\ref{fig4}. It can be seen that there are regions with $\phi_c/T_c > 1$ and at the same time consistent with the experimental measurements of the diphoton Higgs boson decay rates.   

\section{Gravitational Wave Spectrum}\label{sec5}
In this section we briefly review the processes for gravitational waves (GW) production following a first-order phase transition and then represent our result for the ITM extension of the SM. 

There are three main sources for gravitational waves due to an electroweak phase transition: \textbf{1)} The collisions of the bubble walls and subsequent shocks in the plasma \cite{Kosowsky 1992-1, Kosowsky 1992-2, Kosowsky 1993, Kamionkowski 1994, Caprini 2008, Huber 2008, Jinno 2017},  \textbf{2)} The sound waves in the plasma generated after the bubbles have collided but before expansion has dissipated the kinetic energy in the plasma \cite{Giblin 2013, Giblin 2014, Hindmarsh 2014, Hindmarsh 2015}, and \textbf{ 3)} The magneto-hydrodynamical (MHD) turbulence in the plasma formed after the bubbles have collided \cite{Caprini 2006, Kahniashvili 2008, Kahniashvili 2010, Caprini 2009, Kisslinger 2015}. Generically, the three processes happen or coexist and thus the corresponding contributions to the gravitational waves' power spectrum, $\Omega h^2(f)$, must linearly combine at least approximately so that
\begin{equation}
\Omega h^2(f)= \Omega_{col}h^2(f) + \Omega_{sw}h^2(f) + \Omega_{tur}h^2(f)
\end{equation}
where $\Omega_{col}h^2$, $\Omega_{sw}h^2$ and $\Omega_{tur}h^2$ represents the corresponding contribution from bubble collisions, sound waves and turbulence respectively. Now, let us briefly review each of these contributions in detail and estimate the predictions for ITM. 

Analytical studies and also numerical simulations show that in order to estimate the gravitational wave power spectrum due to a specific extension of the SM, one needs to supply at least three parameters \cite{Grojean 2007, Caprini 2016}: i) the ratio of released latent heat from the transition to the energy density of the plasma background, $\alpha$; ii) the time scale of the phase transition, $H_*/\beta$; and, iii) the bubble wall velocity, $v_b$. Using the effective potential and its derivatives at nucleation temperature, $T_n$, the parameter $\alpha$ reads \cite{Caprini 2016} 
\begin{equation}\label{eqn:alpha}
\alpha= \left.\frac{1}{\rho_{R}}\left[-(V_{\rm EW}-V_f)+ T_n \left(\frac{dV_{\rm EW}}{dT} - \frac{dV_f}{dT}\right)\right]\right|_{T=T_n} 
\end{equation}
where $V_f$ is the value of the potential in the unstable vacuum,  $V_{\rm EW}$ is the value of the potential in the final vacuum, and $\rho_R$ is the energy density of radiation bath, $\rho_R=g_*\pi^2 T_n^4/30$. The time scale of the phase transition can be calculated as \cite{Caprini 2016} 
\begin{equation}\label{eqn:beta}
\frac{H_*}{\beta} = \left. \left[T \frac{d}{dT} \left(\frac{S_3(T)}{T} \right) \right]^{-1}\right|_{T=T_n}
\end{equation}
where $S_3(T)$ is the 3D Euclidean action of the critical bubble. The last ingredient from the phase transition is the velocity of the bubble wall $v_b$. The exact calculation of bubble wall velocity is more complicated since one need to consider bubble interaction with the background plasma. However, the bubble wall velocity can be estimated in terms of $\alpha$ as \cite{Steinhardt 1982}
\begin{equation}
v_b=\frac{1/ \sqrt{3}+\sqrt{\alpha ^2+2 \alpha /3}}{1+\alpha }.
\end{equation}
In fact, the above expression for the bubble wall velocity provides only a lower bound on the true wall velocity \cite{Huber 2008}. For some SM extensions, it has been checked that replacing the above approximation even with $v_b=1$, which is more appropriate for a very strong transition, does not significantly modify the results on gravitational wave signals \cite{Beniwal 2017}. For our analyses, we use the above approximation so that the parameters $\alpha$ and $\beta$ are sufficient in order to calculate the GW signals.

Now based on the numerical simulations, the peak frequency of GW generated by bubble collision \cite{Huber 2008}, sound waves \cite{Hindmarsh 2014,Hindmarsh 2015} and Kolmogorov-type turbulence \cite{Caprini 2009} are respectively given by  
\begin{eqnarray}
&f_{\rm sw}= 1.9 \times 10^{-5} \frac{\beta}{H_*} \frac{1}{v_b} \frac{T_n}{100}\left({\frac{g_*}{100}}\right)^{\frac{1}{6}} {\rm Hz }\nonumber\\
&f_{\rm turb}= 2.7  \times 10^{-5} \frac{\beta}{H_*} \frac{1}{v_b} \frac{T_n}{100}\left({\frac{g_*}{100}}\right)^{\frac{1}{6}} {\rm Hz }\nonumber\\
&f_{\rm{col}}= 16.5\times 10^{-6} \frac{0.62}{v_b^2-0.1 v_b+1.8}\frac{\beta}{H_*} \frac{T_n}{100} \left(\frac{g_*}{100}\right)^{\frac{1}{6}} {\rm Hz}\nonumber.
\end{eqnarray}
Here $g_*$ is the number of relativistic degrees of freedom in the plasma at $T_n$. The energy densities of the GW spectrum, corresponding to each of the three contributions, are given by
\begin{widetext}
\begin{equation}
\Omega h^2_{\textrm{col}}(f)=1.67\times 10^{-5}\left(\frac{\beta}{H_*}\right)^{-2}
\frac{0.11 v_b^3}{0.42+v_b^2}
\left(\frac{\kappa_{\rm col} \alpha }{1+\alpha }\right)^2 
\left(\frac{g_*}{100}\right)^{-\frac{1}{3}}
\frac{3.8 \left(f/f_{\rm col}\right)^{2.8}}{1+2.8 \left(f/f_{\rm col}\right)^{3.8}},
\label{col}
\end{equation}
\begin{equation}
\Omega h^2_{\rm sw}(f)=2.65 \times 10^{-6}\left(\frac{\beta}{H_*}\right)^{-1}
\left(\frac{\kappa_{\rm sw} \alpha }{1+\alpha }\right)^2 
\left(\frac{g_*}{100}\right)^{-\frac{1}{3}}
v_b \left(\frac{f}{f_\mathrm{sw}}\right)^3  \left(
   \frac{7}{4 + 3 (f/f_\mathrm{sw})^2 } \right)^{7/2}.
\label{sw}
\end{equation}
\begin{equation}
\Omega h^2_{\rm turb}(f)=3.35 \times 10^{-4}\left(\frac{\beta}{H_*}\right)^{-1}
\left(\frac{ \kappa_{\rm turb} \alpha }{1+\alpha }\right)^{\frac{3}{2}} 
\left(\frac{g_*}{100}\right)^{-\frac{1}{3}}
v_b \frac{(f/f_\text{turb})^3}{[1+(f/f_\text{turb})]^{\frac{11}{3}} ( 1
    + 8 \pi f/h_*) }.
\label{tur}
\end{equation}
\end{widetext}
Here $h_*$ is the Hubble rate at nucleation temperature, $ h_* = 16.5 \, \mu\mathrm{Hz} \left( \frac{T_n}{100 \, \mathrm{GeV}}\right)  \left(g_*/100\right)^{\frac{1}{6}}$ and $\kappa_\text{cool}$, $\kappa_\text{sw}$ and $\kappa_\text{turb}$ are efficiency factors.

The relative importance of each contribution to GW generation is encoded in the efficiency factors. These depend strongly on the dynamical details of the phase transition. In this regard, the velocity of the bubble wall plays a key role. Depending on the velocity of bubble wall, there are two regimes;  when the wall velocity is relativistic or not.    
Moreover, in the relativistic regime, there are two different scenarios. First, whether the bubble wall reaches a terminal velocity (non-runaway scenario) or, second, the bubble wall accelerates without bound (runaway scenario). To calculate the GW spectrum, it is important to know which of the aforementioned  scenarios apply. For this, the critical value $\alpha_{\infty}$ can be used to distinguish between these two scenarios \cite{Caprini 2016, Espinosa 2010}
\begin{equation}
\alpha_{\infty} \simeq \frac{30}{24 \pi^2}\frac{\sum_a c_a \Delta m_a^2}{g_* T_n^2}.
\end{equation}
Here $c_a = n_a/2$ $(c_a = n_a)$ and $n_a$ is the number of degrees of freedom for boson (fermion) species and $ \Delta m^2_a$ is the squared mass difference of particles between two phases at the nucleation temperature.

For non-runaway scenarios, $\alpha<\alpha_{\infty}$, the bubble wall velocity, $v_b$, remains subliminal and the available energy is transformed into fluid motion. So the dominant contributions to GW come from sound waves and MHD turbulence, $ h^2\Omega_{\rm GW} \simeq h^2\Omega_{\rm sw} + h^2\Omega_{\rm turb} $, with the efficiency factors given by \cite{Caprini 2016}
\begin{align}
&\kappa_{\rm col} \simeq  0 \nonumber \\
&\kappa_{\rm sw} =  (1-\epsilon)\kappa\\
&\kappa_{\rm turb} = \epsilon \kappa \nonumber. 
\end{align}
Here $\epsilon \approx 0.05$ and $\kappa$, in the small and large $v_b$ limits, is approximately given by 
\begin{equation}
\label{eq:kappav}
\kappa \simeq 
\left\{\begin{array}{c c}
\alpha \left(0.73+0.083\sqrt{\alpha}+\alpha\right)^{-1} & v_b \sim 1\\
v_b^{6/5}6.9\, \alpha \left(1.36-0.037 \sqrt{\alpha}+\alpha \right)^{-1},& v_b 
\lesssim 0.1. 
\end{array}
\right.~
\end{equation}
The full expressions for $\kappa$ are given in Ref.\cite{Espinosa 2010}.

For runaway scenario, $\alpha>\alpha_{\infty}$, the excess vacuum energy density
leads to bubble acceleration and $v_b$ is bounded only by the speed of light, $v_b = 1$. In this case, all the three GW sources contribute with efficiency factors 
\begin{align}
&\kappa_{\rm col} = 1-\frac{\alpha_{\infty}}{\alpha}\nonumber \\
&\kappa_{\rm sw} = (1-\epsilon)\kappa \\
&\kappa_{\rm turb} = \epsilon \kappa \nonumber
\end{align}
where in this case $\kappa$ is given by \cite{Espinosa 2010, Beniwal 2017}
\begin{equation} \label{kappa1}
\kappa=\frac{\alpha_{\infty}}{\alpha}\left( \frac{\alpha_{\infty}}{0.73+0.083\sqrt{\alpha_{\infty}}+\alpha_{\infty}} \right).
\end{equation}

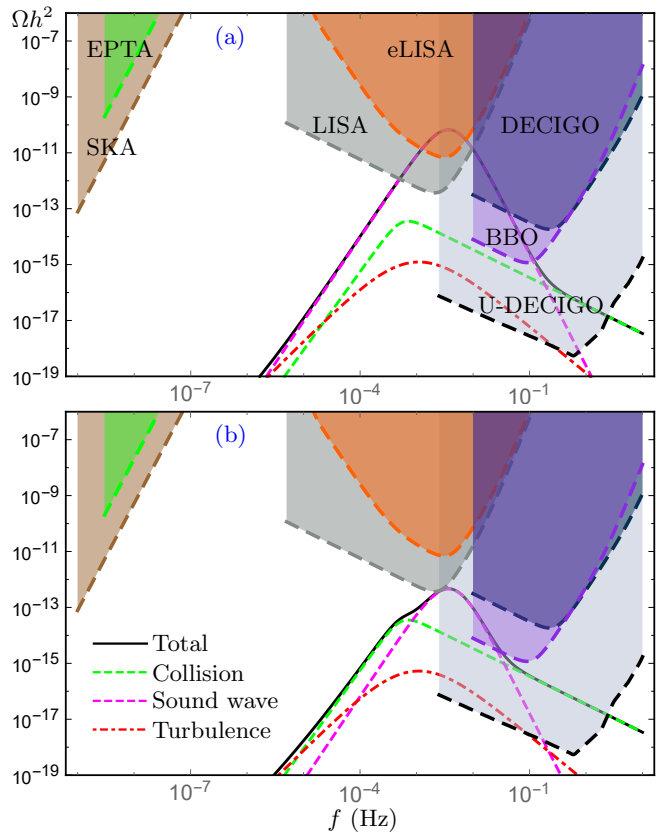
\begin{figure}[H]
\centering
\begin{tikzpicture}
\draw (0,0) node[above right]{\includegraphics[width=1\linewidth, trim={0cm 0cm 0cm 0cm}]{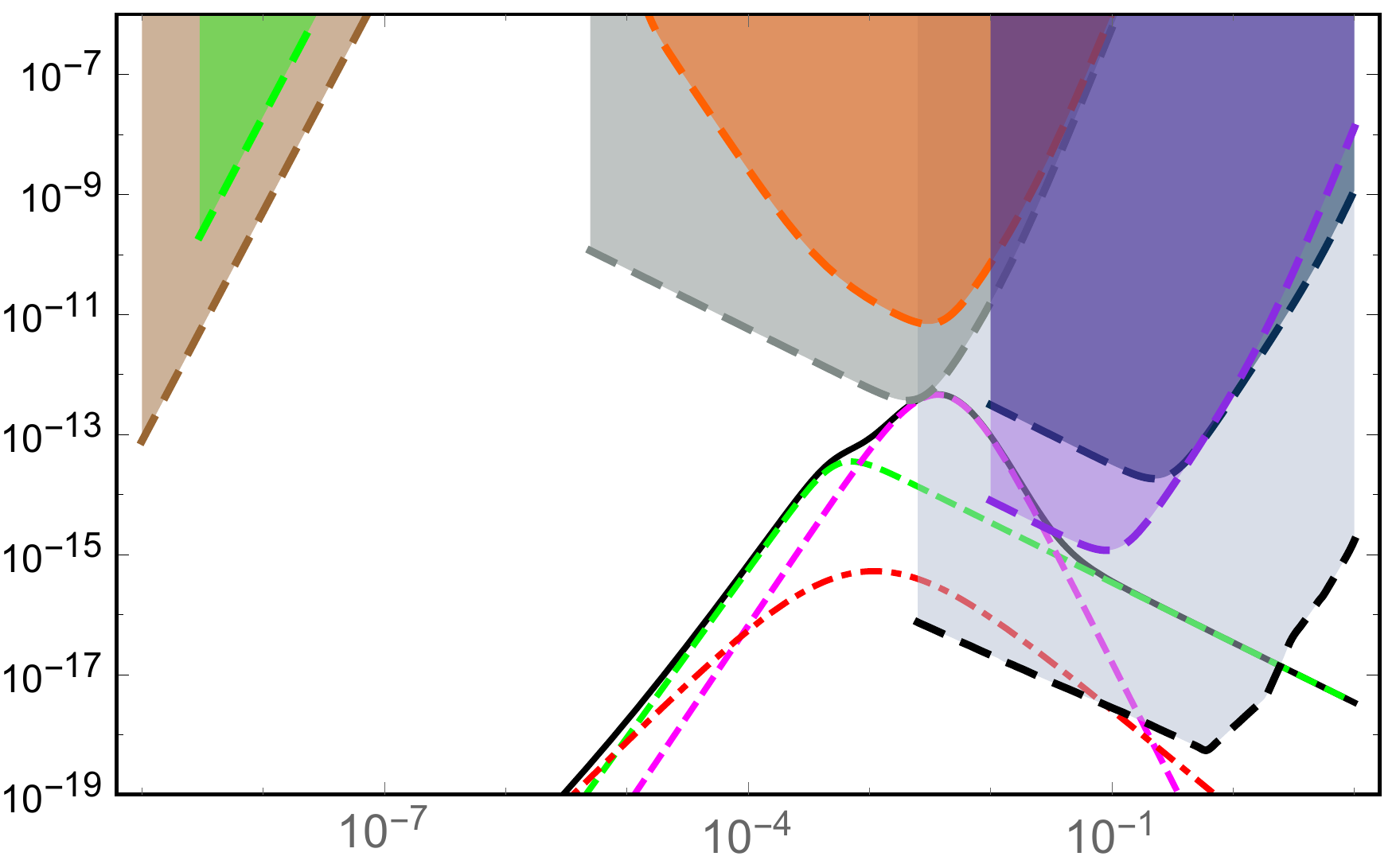}};
\draw (0,5.3) node[above right]{\includegraphics[width=1\linewidth, trim={0cm 0cm 0cm 0cm}]{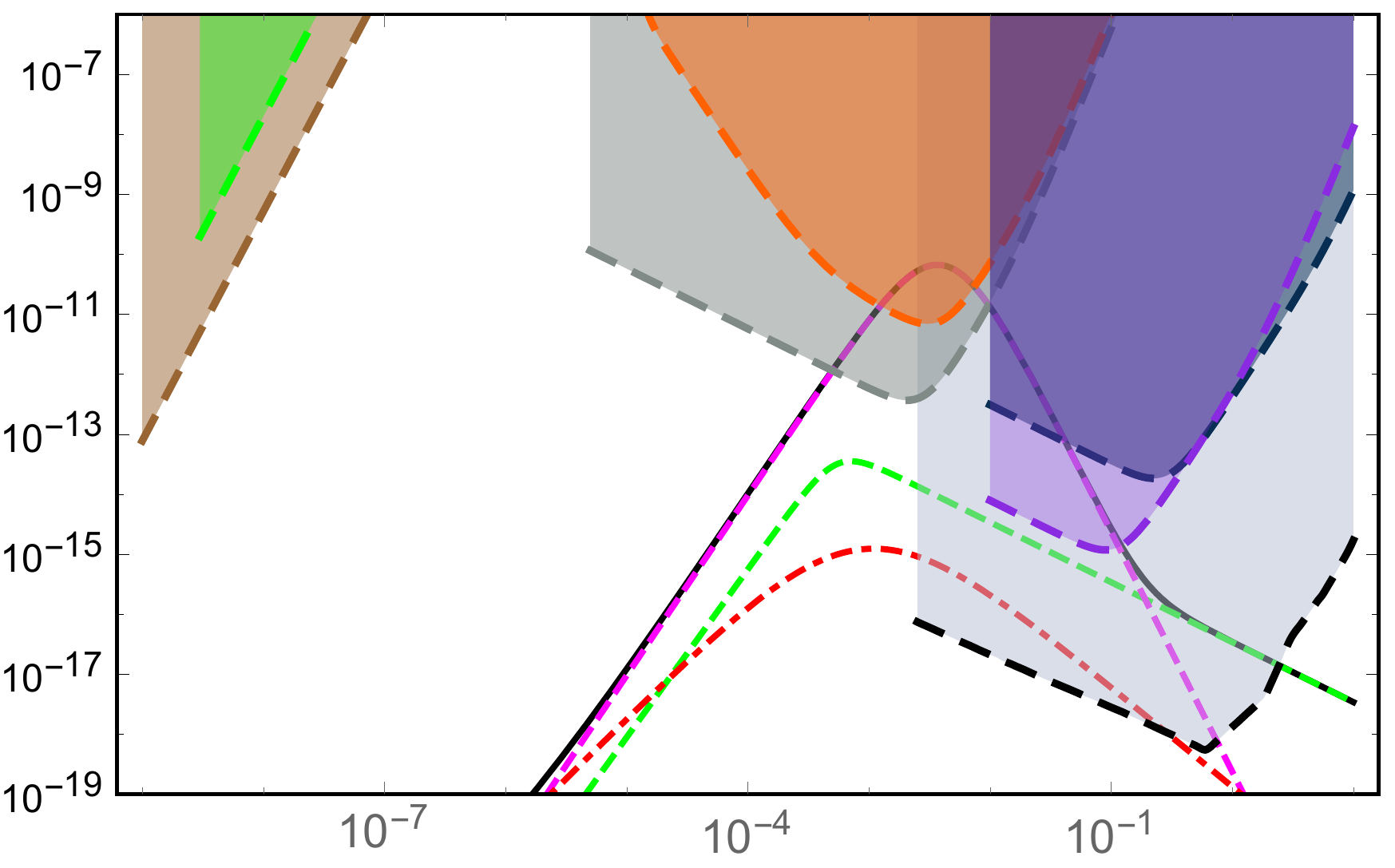}};
\draw (0.7,0.8) node[above right]{\includegraphics[width=.17\linewidth, trim={0cm 0cm 0cm 0cm}]{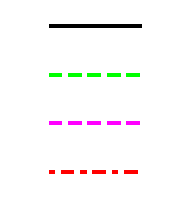}};
\draw (1.87,2.1) node[above right]{\small{Total}};
\draw (1.87,1.7) node[above right]{\small{Collision}};
\draw (1.87,1.35) node[above right]{\small{Sound wave}};
\draw (1.87,0.95) node[above right]{\small{Turbulence}};
\draw (0,10.4) node[above right]{$\Omega h^2$};
\draw (4.2,-.3) node[above right]{$f$ (Hz)};
\draw (6.5,9) node[above right]{ِDECIGO};
\draw (5,10) node[above right]{ِeLISA};
\draw (4,9) node[above right]{ِLISA};
\draw (6.3,7.5) node[above right]{BBO};
\draw (6.2,6.6) node[above right]{ِU-DECIGO};
\draw (1,10) node[above right]{ِEPTA};
\draw (1,8.7) node[above right]{ِSKA};
\draw (2.7,10.1) node[above right]{ِ\color{blue}{\small{(a)}}};
\draw (2.7,4.8) node[above right]{ِ\color{blue}{\small{(b)}}};
\end{tikzpicture}
\caption{Spectra of the GW from the electroweak phase transition for a typical point in the parameter space; $m_{\Delta^{++}}=335.3$ GeV, $m_S=426$ GeV and $\mu_{\Delta}=100$ GeV. The shaded regions represent the expected sensitivities of GW
interferometers. Panel(a): The GW spectrum is computed using the semi-analytical hydrodynamic approximation, eq.(\ref{col}) to eq.(\ref{tur}). Panel(b): Same as in pane(a) but with the corrected $\kappa$-parameter and modification due to the finite lifetime of the sound waves applied.}\label{ffig5} 
\end{figure}

For more accuracy in the calculation of GW spectra, we consider two corrections which were discovered in recent studies \cite{Cutting 2020, Guo 2020}. Firstly, we consider the correction of efficiency factors for strong transitions. The values of $\kappa$ given above, are from a semi-analytical hydrodynamic analysis. These are good estimations of $\kappa$ only for relatively weak transitions with $\alpha \ll 1$. For strong transitions and small $v_b$, a recent simulation found that $\kappa$ as specified in eq.(\ref{kappa1}) gives an overestimation \cite{Cutting 2020}. Using the numerical results in \cite{Cutting 2020}, we refine the estimation of the efficiency factor \cite{footnote-GW}. Secondly, we consider an additional suppression factor $\Upsilon$ in the $\Omega h^2_{\rm sw}$, which originates from the finite lifetime, $\tau_{sw}$, of the sound waves \cite{Guo 2020}
\begin{equation}
\Upsilon=1-\frac{1}{\sqrt{1+2\tau_{sw}H_*}}. 
\end{equation}
For the classical approach, $\tau_{sw}\to \infty$ is usually assumed and that corresponds to the asymptotic $\Upsilon \to 1$. The lifetime $\tau_{sw}$ can be considered as the time scale when the turbulence develops, approximately given by \cite{Pen 2016, Hindmarsh 2017}
$$\tau_{sw}\sim\frac{R_*}{\bar{U}_f}$$
where $R_*$ is the mean bubble separation and is related to $\beta$ through the relation $R_*=(8\pi)^{1/3} v_b/\beta$ for an exponential bubble nucleation \cite{Hindmarsh 2019}. Further, the analysis performed in Ref. \cite{Hindmarsh 2019} was based on Minkowski spacetime. For an analysis based on an expanding universe, see \cite{Guo 2020}. The denominator $\bar{U}_f$ is the root-mean-squared fluid velocity which can be obtained from hydrodynamic analyses as $\bar{U}_f=\sqrt{(3\kappa_\nu \alpha/4)}$ \cite{Hindmarsh 2019, Weir 2018}. In Fig.\ref{ffig5}, we explicitly show the effect of these corrections for our analyses for a typical point the ITM parameters space.

\begin{widetext}

\begin{figure}[H]
\centering
\begin{tikzpicture}
\draw (-2.1,.2) node[above right]{\includegraphics[width=.46\linewidth, trim={0cm 0cm 0cm 0cm}]{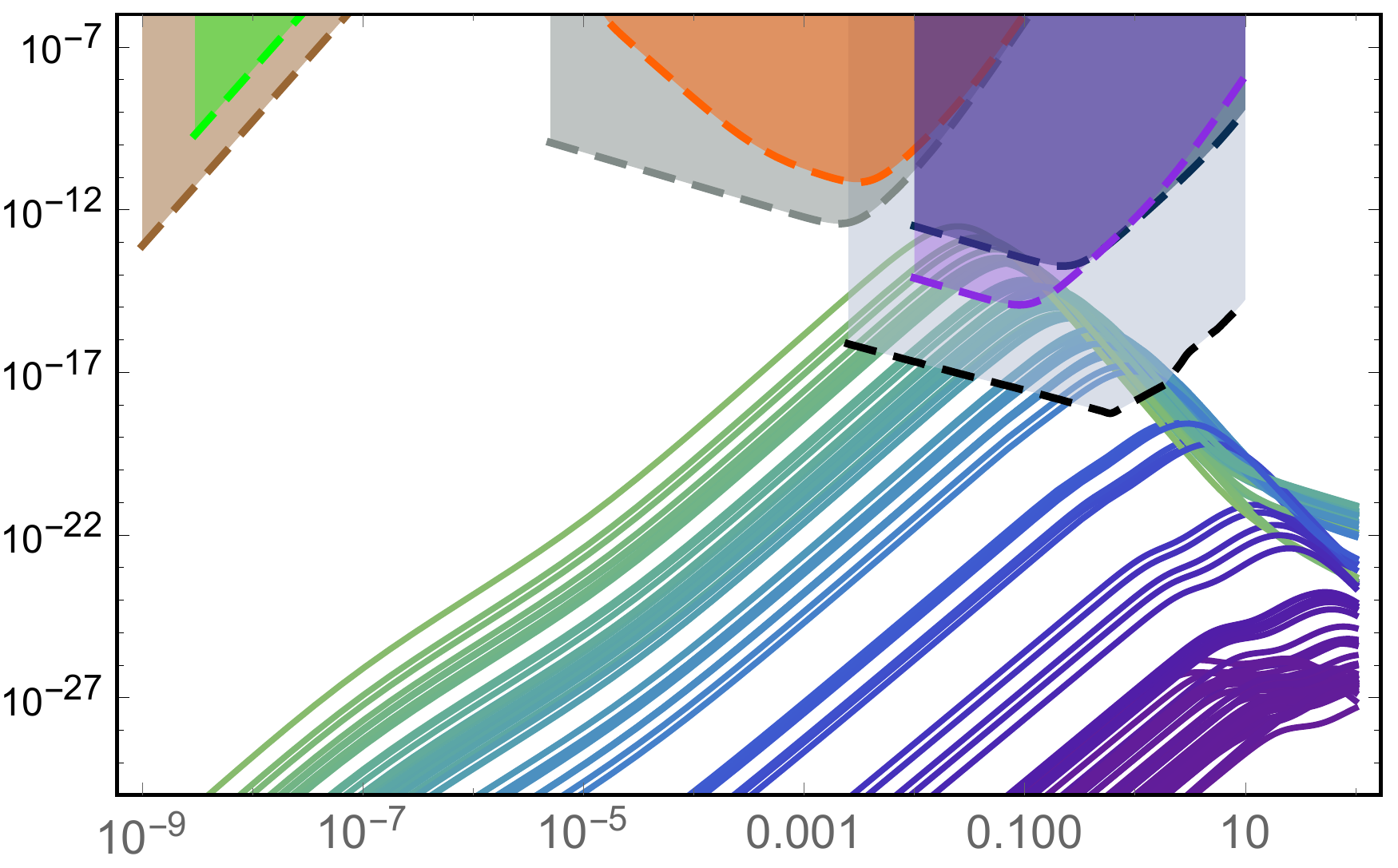}};
\draw (6.4,.2) node[above right]{\includegraphics[width=.46\linewidth, trim={0cm 0cm 0cm 0cm}]{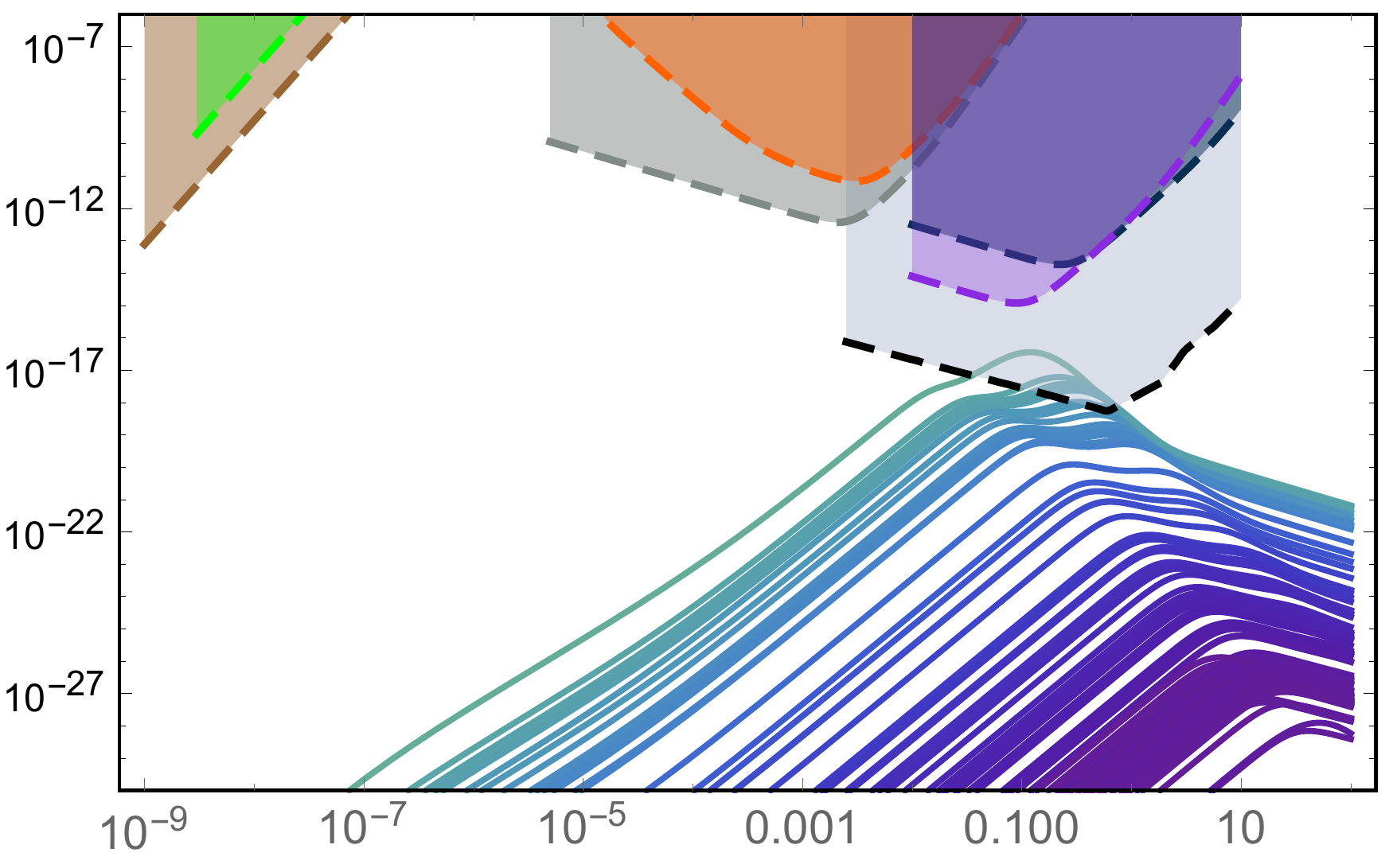}};
\draw (-2.1,6) node[above right]{\includegraphics[width=.46\linewidth, trim={0cm 0cm 0cm 0cm}]{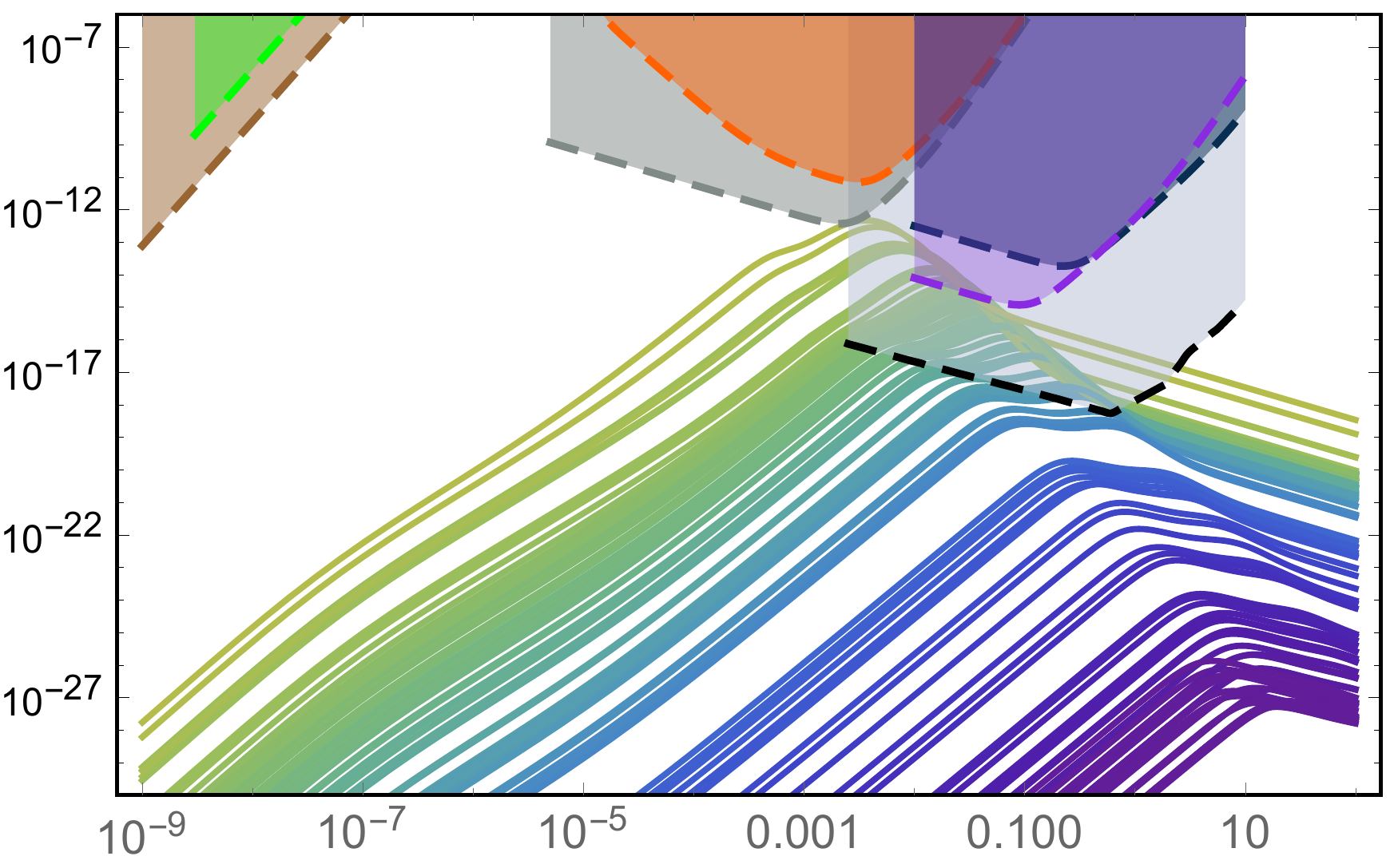}};
\draw (6.4,6) node[above right]{\includegraphics[width=.46\linewidth, trim={0cm 0cm 0cm 0cm}]{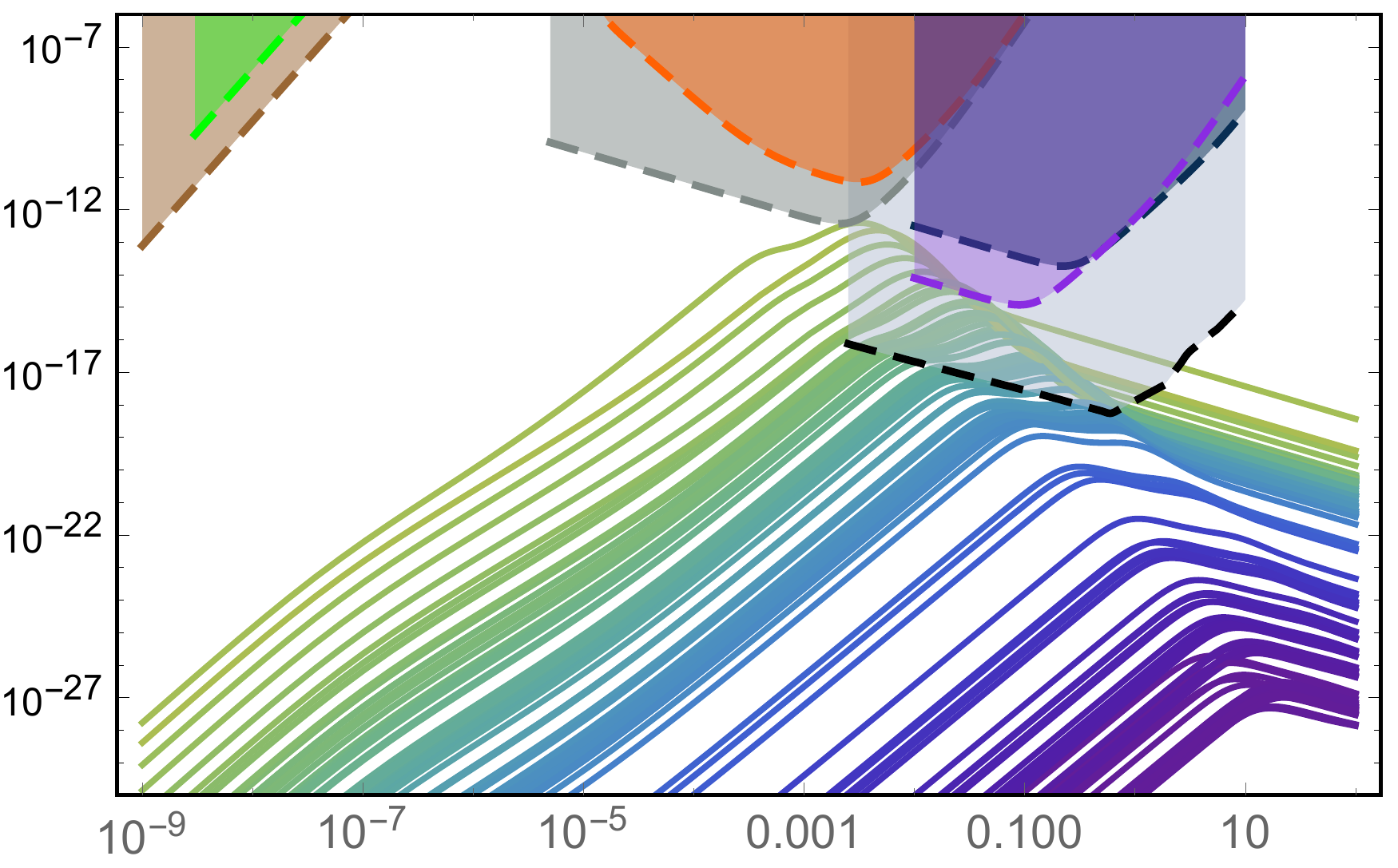}};
\draw (14.5,5.91) node[above right]{\includegraphics[width=.042\linewidth, trim={0cm 0cm 0cm 0cm}]{BarLegend.pdf}};
\draw (14.5,.1) node[above right]{\includegraphics[width=.042\linewidth, trim={0cm 0cm 0cm 0cm}]{BarLegend.pdf}};
\draw (-1.2,11.1) node[above right]{$\mu_\Delta=100$ (GeV)};
\draw (7.3,11.1) node[above right]{$\mu_\Delta=150$ (GeV)};
\draw (-1.2,5.3) node[above right]{$\mu_\Delta=200$ (GeV)};
\draw (7.3,5.3) node[above right]{$\mu_\Delta=250$ (GeV)};
\draw (14.4,5.4) node[above right]{$\phi_c/T_c$};
\draw (14.4,11.2) node[above right]{$\phi_c/T_c$};
\draw (-2.6,2.5) node[above right]{$\Omega h^2(f)$};
\draw (-2.6,8.3) node[above right]{$\Omega h^2(f)$};
\draw (1.7,-.2) node[above right]{$f$ (Hz)};
\draw (10.2,-.2) node[above right]{$f$ (Hz)};
\draw (12.1,4.5) node[above right]{ِDECIGO};
\draw (11.7,3.6) node[above right]{BBO};
\draw (10.7,4.7) node[above right]{ِeLISA};
\draw (9.85,4.43) node[above right]{ِLISA};
\draw (11.8,3.18) node[above right]{ِU-DECIGO};
\draw (7.2,4.9) node[above right]{ِEPTA};
\draw (7.2,4.2) node[above right]{ِSKA};
\end{tikzpicture}
\caption{Spectra of GW from the electroweak phase transition for randomly sampled examples from the coloured points in Fig.\ref{fig4}, i.e. the points with strong first-order EWPT. The sensitivity region for prospective GW detectors such as eLISA, BBO and DECIGO are also shown. It can be seen that the intensity of GW signal increases with the strength of the phase transition, i.e. $\phi_c/T_c$. For comparison we also show the sensitivity regions for SKA and EPTA detectors which cannot probe any part of the parameters space of inert complex triplet model.}\label{ffig1}
\end{figure}

\end{widetext}

Considering the above corrections, we show in Fig.\ref{ffig1} the GW power spectra for selected points with various $\phi_c/T_c$ values. We find that the peak frequencies and strengths of the gravitational wave signals are strongly correlated with the strength of the phase transition. 

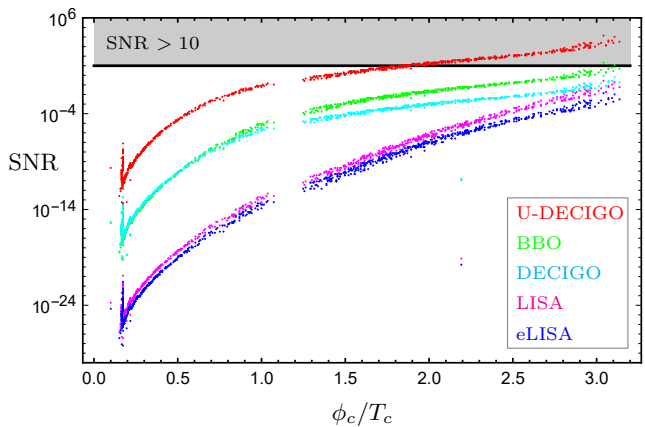
\begin{figure}[H]
\centering
\begin{tikzpicture}
\draw (0,0) node[above right]{\includegraphics[width=0.95\linewidth, trim={0cm 0cm 0cm 0cm}]{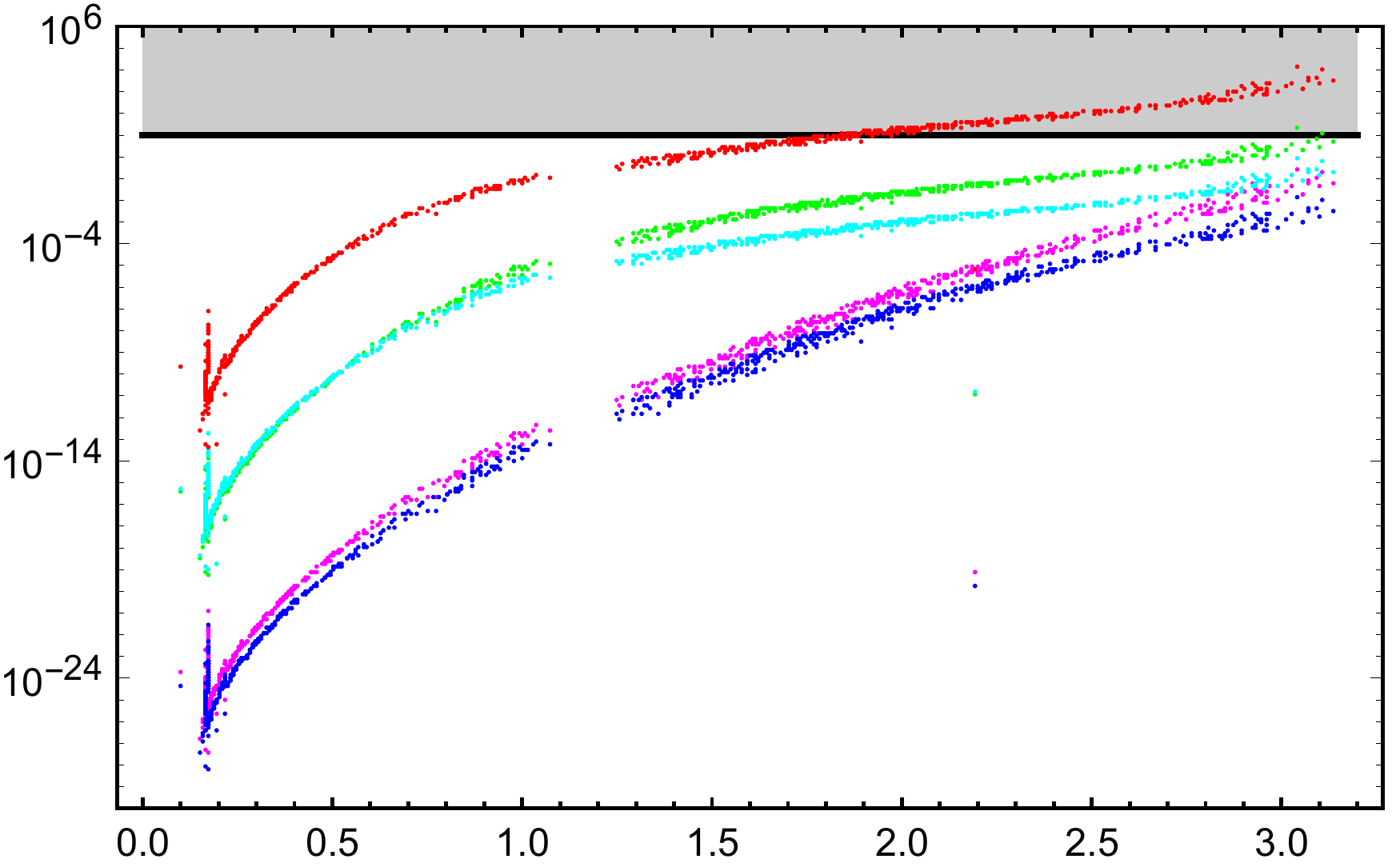}};
\draw (1,4.5) node[above right]{\scriptsize{SNR $>10$ }};
\draw (-.3,2.9) node[above right]{SNR};
\draw (4,-.5) node[above right]{$\phi_c/T_c$};
\draw (6.45,2.25) node[above right]{\color{red}{\scriptsize{U-DECIGO}}};
\draw (6.45,1.85) node[above right]{\color{green}{\scriptsize{BBO}}};
\draw (6.45,1.45) node[above right]{\color{cyan}{\scriptsize{DECIGO}}};
\draw (6.45,1.05) node[above right]{\color{magenta}{\scriptsize{LISA}}};
\draw (6.45,0.65) node[above right]{\color{blue}{\scriptsize{eLISA}}};
\draw[gray] (6.45,0.65) rectangle (8.05,2.65);
\end{tikzpicture}
\caption{The scatter plot of signal-to-noise ratio versus $\phi_c/T_c$ for the ITM predicted GW signals based on different gravitational wave detectors.}\label{fig7}
\end{figure}

Next, in order to assess the detectability of the GW signal by a given detector, one needs to consider the signal-to-noise ratio
(SNR) over the running time of the detector, $t_{obs}$, which is given by \cite{eLISA 2016,Schmitz:2020syl},
\begin{equation}
\mathrm{SNR}=\sqrt{ \delta \ t_{obs} \int_{f_{min}}^{f_{max}}  \left[\frac{h^2\Omega_{\mathrm{GW}}(f)}{h^2\Omega_{\mathrm{Sens}}(f)}\right]^{2}df},
\end{equation}
where  $h^2\Omega_{\mathrm{Sens}}(f)$ represents the sensitivity of the  detector. The interval of integration, $[f_{min},f_{max}]$, is the frequency bandwidth of the detector. The factor $\delta$, which indicate the number of independent channels for the GWs detector, is equal to 2 for BBO and U-DECIGO, and is equal to 1 for the rest. We consider $t_{obs}=5$  year, for all the detectors.  Whenever SNR turns out to be larger than some threshold value, $\mathrm{SNR}> \mathrm{SNR}_{thr}$, then one can assert that the experiment under consideration will be able to detect the GW signal. The method of quantifying $\mathrm{SNR}_{thr}$ is briefly described in \cite{eLISA 2016}. 
For example, the SNR threshold for discovery at eLISA is 10 or 50, depending on the operating configuration \cite{eLISA 2016}.  Here we compute the signal-to-noise ratio for the eLISA, LISA, BBO, DECIGO and U-DECIGO detectors. The results are shown in Fig \ref{fig7}.

Based on the results peresented in Fig.\ref{fig7}, the computed SNRs of eLISA are less than its threshold, 10, which means these gravitational waves are not  detectable by eLISA. The biggest SNRs are  associated with U-DECIGO. In fact, for points with $\phi_c/T_c\sim 3$,  calculatons concerning the U-DECIGO leads to a $\textrm{SNR} \sim 100$. However, since the  $\mathrm{SNR}_{thr}$ of U-DECIGO is not known (to the best of our knowledge), the question remains whether it can detect such GW arising from EWPT with  $\phi_c/T_c\sim 3$ or not.

\section{Summary}
We have investigated the cosmological electroweak phase transition in an inert triplet scalar extension of the SM model. We found that there are regions of parameter space which can yield a strong first-order electroweak phase transition and at the same time consistent with recent LHC results on Higgs to diphoton decay rate.

In principle, a first order cosmological phase transition can lead to a background stochastic GW. This, besides collider phenomenology, can be used to probe the parameter space of particle physics model beyond the SM. In this regard, considering the recent treatment of GW spectrum estimation \cite{Cutting 2020, Guo 2020}, we study the GW signals generated after the first order electroweak phase transitions within the framework of the inert triplet model.

Based on the signal-to-noise ratio analyses, we have found that very sensitive
	GW detectors will be needed for detecting the inert triplet model signal. Probing the GW signals of this model will be difficult or maybe impossible for prospective space-based GW detectors with less sensitive configurations compared to U-DECIGO. 

We also compute the $ H\to Z \gamma$ decay rate in this model, which can be used as probe, at future collider experiments, such as High Luminosity LHC and other colliders with higher center-of-mass energies \cite{Goertz 2020}.

\section*{Acknowledgements}
Thanks very much to H. Hashamipour, A. Kargaran and L. Kalhor for conversations, useful discussions, or comments while working on this project. We also sincerely thank Referee for its insightful comments towards the improvement of this paper.

\section*{Appendix A}\label{AppendixA}
For completeness, below we summarise the decay widths of the Higgs boson, which are same in both SM and ITM.
\\
\textit{Decay to leptons---.} In the Born approximation, the partial decay width of $h$ to any fermion channel is \cite{Gunion 2000, Rizzo 1980}
$$
\Gamma_{h\to f\bar{f}}=\frac{N_c g^2}{32 \pi m_W^2}m_h m_{f}^2 \bigg(1-\frac{4m_{f}^2}{m_h^2}\bigg)^{3/2},
$$
where $N_c$, the color factor, is 1 for leptons and 3 for quarks.

\textit{Decay to quarks---.} \cite{Djouadi 2008}
\begin{align}
\Gamma(h\to q\bar{q})=&\frac{N_c g^2}{32 \pi m_W^2}m_h \overline{m}_q^2(m_h)\Bigg\{1+5.67\frac{\overline{\alpha}_s(m_h)}{\pi}+\nonumber\\*
&+\bigg[37.51-1.36 N_f-\frac{2}{3}\log\frac{m_h^2}{m_t^2}+\nonumber\\*
&+\bigg(\frac{1}{3}\log\frac{\overline{m}_q^2(m_h)}{m_h^2}\bigg)^2\bigg]\frac{\overline{\alpha}_s^2(m_h)}{\pi^2}\Bigg\},\nonumber
\end{align}
where, at the one-loop level, the running strong coupling constant is approximated  \cite{Beringer 2012}
$$
\overline{\alpha}_s(m_h)=\frac{\overline{\alpha}_s(m_Z)}{1+\frac{33-2N_f}{12\pi}\overline{\alpha}_s(m_Z)\log\frac{m_h^2}{m_Z^2}},
$$
$N_f=5$, and the running quark mass defined at the scale $m_h$ is \cite{Djouadi 1996}
\begin{align}
\overline{m}_q(m_h)=&\overline{m}_q(m_q)\bigg(\frac{\overline{\alpha}_s(m_h)}{\overline{\alpha}_s(m_q)}\bigg)^{12/(33-2N_f)}\times\nonumber\\*
&\times\frac{1+c_{1q}\overline{\alpha}_s(m_h)/\pi+c_{2q}\overline{\alpha}_s^2(m_h)/\pi^2}{1+c_{1q}\overline{\alpha}_s(m_q)/\pi+c_{2q}\overline{\alpha}_s^2(m_q)/\pi^2},\nonumber
\end{align}
in which $c_{1b}=1.17$, $c_{2b}=1.50$ and $c_{1c}=1.01$, $c_{2c}=1.39$ for the bottom and the charm quark, respectively.
The strong coupling and the quark masses are taken from ~\cite{Beringer 2012}: $\overline{\alpha}_s(M_Z)=0.118$, $\overline{\alpha}_s(m_b)=0.223$, $\overline{\alpha}_s(m_c)=0.38$, $\overline{m}_c(m_c)=1.273$ GeV,  and $\overline{m}_b(m_b)=4.18$ GeV.

\textit{Decay to gauge bosons---.} The decay width of Higgs decay to gluon, via quark loops, is given by \cite{Gunion 2000}  
$$
\Gamma_{h\to gg}=\frac{\alpha_s^2 g^2 m_h^3}{128\pi^3 m_W^2}\bigg|\sum_{i}\frac{1}{2}A_{1/2}\bigg(\frac{4m_i^2}{m_h^2}\bigg)\bigg|^2,
$$
where the sum is over all quarks, $i=(t, b, c, s, u, d)$, however, the main contribution comes from top quark. The loop function, $A_{1/2}(x)$, is represented in appendix B. Finally, the decay widths of $h \to ZZ^*$ and $h \to WW^*$, summed over all available channels, are given by \cite{Gunion 2000, Keung 1984},
$$
\Gamma_{h\to VV^*}=\frac{3 g^4 }{512\pi^3}m_h\delta_VR_T(\frac{m_V^2}{m_h^2}),
$$
where 
\begin{align}
R_T(x)=&\frac{3(1-8x+20x^2)}{\sqrt{4x-1}}\arccos\Big(\frac{3x-1}{2x^{3/2}}\Big)\nonumber\\*
&-\frac{|1-x|}{2x}\big(2-13x+47x^2\big)\nonumber\\*
&-\frac{3}{2}(1-6x+4x^2)\log x\nonumber
\end{align}
and $\delta_W=1$, $\delta_Z=\frac{1}{\cos^4 \theta_W}(\frac{7}{12}-\frac{10}{9}\sin^2\theta_W+\frac{40}{27}\sin^4\theta_W)$. 
\section*{Appendix B}
The loop functions $A^{\gamma\gamma}_{(0,\,1/2,\,1)}$ and $A^{Z\gamma}_{(0,\,1/2,\,1)}$ are defined as follows \cite{Gunion 2000, Spira 1998}:

\begin{eqnarray}
A_0^{\gamma\gamma} (x) &=& -x^2[x^{-1}-f(x)]\,,
\nonumber\\
A_{1/2}^{\gamma\gamma} (x) &=& 2x^2[x^{-1}+(x^{-1}-1)f(x)]\,,
\nonumber\\
A_1^{\gamma\gamma}(x) &=& -x^2[2x^{-2}+3x^{-1}+3(2x^{-1}-1)f(x)]\,,
\nonumber\\
A_0^{Z\gamma}(x,y) &=& I_1(x,y)\,, 
\nonumber\\
A_{1/2}^{Z\gamma} (x,y) &=& I_1(x,y)-I_2(x,y)\,,
\nonumber\\
A_1^{Z\gamma}(x,y) &=& 4(3-\tan^2\theta_W)I_2(x,y)\nonumber\\
&+&[(1+2x^{-1})\tan^2 \theta_W-(5+2x^{-1})]I_1(x,y)\,, \nonumber
\end{eqnarray}
where
\begin{eqnarray}
I_1(x,y) &=& \frac{x y}{2(x-y)}+\frac{x^2 y^2}{2(x-y)^2}[f(x)-f(y)]\nonumber\\
&+&\frac{x^2 y}{(x-y)^2}[g(x)-g(y)]\,,
\nonumber\\
I_2(x,y) &=& -\frac{x y}{2(x-y)}[f(x)-f(y)]\;,\nonumber
\end{eqnarray}
with the functions $f(x)$ and $g(x)$ are given by
\begin{displaymath}
f(\tau)=\begin{cases}
\arcsin^2\big(\frac{1}{ \sqrt{\tau} }\big) & \textrm{for } \tau\geqslant 1,\\[4pt]
-\frac{1}{4}\Big[\log\Big(\frac{1+\sqrt{1-\tau}}{1-\sqrt{1-\tau}}\Big)-i\pi\Big]^2  & \textrm{for } \tau<1.
\end{cases}
\end{displaymath}
$$
g(\tau)=\begin{cases}
\sqrt{\tau-1}\arcsin(1/\sqrt{\tau})& \textrm{for }\tau\geqslant 1,\\[4pt]
\frac{1}{2}\sqrt{1-\tau}\left(\log\frac{1+\sqrt{1-\tau}}{1-\sqrt{1-\tau}}-i\pi\right)&\textrm{for } \tau<1.
\end{cases}
$$



\begin{thebibliography}{99}
\bibitem{footnote-GW}
To this aim, we use of an interpolation of the numerical results which are represented in the Fig. 3 of the Ref. \cite{Cutting 2020}.
\end{thebibliography}
\end{document}